\documentclass[prd,onecolumn,tightenlines,preprintnumbers,showpacs,superscriptaddress,notitlepage,nofootinbib,eqsecnum,floatfix,longbibliography]{revtex4-1}

\usepackage[normalem]{ulem}
\usepackage{amsmath}           
\usepackage{amssymb}           
\usepackage{xcolor}             
\usepackage{bbm}               
\usepackage{braket}            
\usepackage{xspace}            
\usepackage{mathtools}         
\usepackage{graphicx}          
\usepackage{enumerate}         
\usepackage{hyperref}          
\usepackage{tabularx}  		   
\usepackage{multirow}		   
\usepackage{enumitem}
\usepackage{bm}

\usepackage{etoolbox}
\makeatletter
\patchcmd{\subsubsection}{\itshape}{\itshape\bfseries}{}{}
\makeatother

\newcommand{\Ac}{\mathcal{A}}

 \newcommand{\Jc}{\mathcal{J}}
\newcommand{\Kc}{\mathcal{K}}
\newcommand{\Lc}{\mathcal{L}}
\newcommand{\Mc}{\mathcal{M}}
\newcommand{\Mcs}{\Mc_{2;s}}
\newcommand{\Kcs}{\Kc_{2;s}}
\newcommand{\Mcsoff}{\Mc^{({\rm off})}_{2;s}}
\newcommand{\Mchoff}{\Mc^{({\rm off})}_{2;h}}

\newcommand{\Kcsoff}{\Kc^{({\rm off})}_{2;s}}
\newcommand{\Mch}{\Mc_{2;h}}

\newcommand{\Oc}{\mathcal{O}}

\newcommand{\Rc}{\mathcal{R}}
\newcommand{\Tc}{\mathcal{T}}

\newcommand{\nn}{\nonumber}

\newcommand\identity{1\kern-0.25em\text{l}}

\newcolumntype{Y}{>{\centering\arraybackslash}X}
\newcolumntype{s}{>{\centering\arraybackslash\hsize=.2\hsize}X}

\definecolor{wm_green}{HTML}{115740}
\definecolor{wm_gold}{HTML}{B9975B}

\usepackage{mfirstuc} 
\newcommand{\addReviewer}[2]{
  \expandafter\newcommand\csname #1\endcsname[1]{{\sf \color{#2} {#1}:\,##1}}
  \expandafter\newcommand\csname #1cor\endcsname[2]{{\color{#2} {#1}:\,\st{##1}{\sf ##2}}}
  \expandafter\newcommand\csname #1color\endcsname{#2}
}

\usepackage{soul,color}
\definecolor{chromeyellow}{rgb}{1.0, 0.65, 0.0}
\definecolor{DodgeBlue}{rgb}{0.118, 0.565,1.000}
\definecolor{asparagus}{rgb}{0.53, 0.66, 0.42}
\definecolor{cadmiumgreen}{rgb}{0.0, 0.42, 0.24}

\definecolor{jlab_red}{RGB}{192,39,45}
\definecolor{jlab_orange}{RGB}{249,102,0}
\definecolor{jlab_blue}{RGB}{47,122,121}
\definecolor{jlab_green}{RGB}{65,125,10}
\definecolor{bobcat_green}{RGB}{2,66,48}

\addReviewer{cc}{jlab_green}
\addReviewer{rb}{jlab_blue}
\addReviewer{hwh}{jlab_orange}
\addReviewer{drp}{bobcat_green}

\hypersetup{
pdftitle     = {Partial-wave projection of the three-body scattering amplitudes},
pdfsubject   = {Scattering Theory},
pdfkeywords  = {Scattering, Partial Waves, QCD, Hadron, Physics, Lattice},
pdfnewwindow = {true},
colorlinks   = {true},
linkcolor    = {blue},
citecolor    = {blue},
filecolor    = {blue},
urlcolor     = {blue}
} 

\newcommand{\ohio}{Department of Physics and Astronomy and Institute of Nuclear and Particle Physics, Ohio University, Athens, Ohio 45701, USA}
\newcommand{\darmstadt}{Technische Universit\"at Darmstadt, Department of Physics, 64289 Darmstadt, Germany}
\newcommand{\HFHF}{
ExtreMe Matter Institute EMMI and Helmholtz Forschungsakademie Hessen für FAIR (HFHF), GSI Helmholtzzentrum für Schwerionenforschung GmbH, 64291 Darmstadt, Germany}
\newcommand{\ucb}{Department of Physics, 
University of California, 
Berkeley, CA 94720, USA}   
\newcommand{\lbnl}{Nuclear Science Division, 
Lawrence Berkeley National Laboratory, Berkeley, 
CA 94720, USA}

\usepackage{orcidlink}

\begin{document}  
\preprint{INT-PUB-26-019}
\title{
Factorization in quasi-free breakup reactions with two bodies left behind 
}


\author{Ra\'ul A. Brice\~no~\orcidlink{0000-0003-1109-1473}}
\email[e-mail: ]{rbriceno@berkeley.edu}
\affiliation{\ucb}
\affiliation{\lbnl}

\author{Caroline S. R. Costa~\orcidlink{0000-0003-1392-837X}}
\email[e-mail: ]{costa@berkeley.edu}
\affiliation{\ucb}
\affiliation{\lbnl}

\author{Hans-Werner Hammer~\orcidlink{0000-0002-2318-0644}}
\email[e-mail: ]{Hans-Werner.Hammer@physik.tu-darmstadt.de}
\affiliation{\darmstadt}
\affiliation{\HFHF}

\author{Daniel R. Phillips~\orcidlink{0000-0003-1596-9087}}
\email[e-mail: ]{phillid1@ohio.edu}
\affiliation{\ohio}

\begin{abstract}

We consider scenarios in which a shallow bound state undergoes breakup by a probe whose energy is high compared to the binding energy. The first two scenarios, which serve as warm-up exercises, involve a single heavy particle bound to a light particle, analogous to a core nucleus bound to a neutron. We show that in quasi-free kinematics, the leading effect comes from the heavy particle being knocked out by the probe, with corrections suppressed by inverse powers of the probe momentum. This formally justifies extracting neutron form factors from high-energy deuteron breakup in quasi-free kinematics. In Scenario 1, the probe is a local current; in Scenario 2, it is hadron scattering. In Scenarios 3 and 4 we consider, respectively, a local current and hadron scattering, but now on a three-body bound state of a heavy particle and two light particles. Hard knockout of the heavy particle leaves two low-energy particles behind, which can interact with one another. In all four scenarios, we prove that the amplitude is dominated by the nearby on-shell pole of the heavy-particle propagator and derive a closed-form expression for this contribution. When two bodies are left behind, the leading amplitude is the product of the scattering of the two light particles, a dynamical function depending on the probe, and a real function related to the bound-state wavefunction. Thus, quasi-free removal of a core nucleus from a system with halo neutrons provides access to on-shell data on multi-neutron interactions. The resulting amplitudes are relativistic and satisfy unitarity for the remnant subsystem exactly. We also provide complementary non-relativistic derivations. While the derivations are for spinless particles, the generalization to spin is straightforward, since the results depend only on quasi-free knockout kinematics; we make no assumptions about the inter-particle dynamics.

\end{abstract}
\date{\today}
\maketitle

\section{Introduction}
\label{sec:intro}

Quasi-free nuclear reactions involving hard-particle knockout are a primary tool for investigating 
nuclear structure of stable and unstable nuclei \cite{JACOB:1966wce, Jacob:1973tu, Mougey:1976sc, Jans:1982aw, Aumann:2013hga, Atar:2018dhg, Chen:2019gau, Panin:2021gds,Duer:2022ehf, Yoshida:2024xzx}. 
In such processes, an incoming high-energy probe interacts primarily with a single nucleon inside a nucleus, knocking it out on a short enough time scale that the rest of the nucleus can be regarded as a spectator to the interaction. At radioactive beam facilities, the experiments are carried out in inverse kinematics where a fast beam of unstable nuclei impinges on the target.
If the probe three-momentum is approximately collinear with the final-state momentum of the struck particle, and the particle is not deeply bound in the initial state, then the scattering takes place in approximately on-shell kinematics, and is called a quasi-free reaction.
Hard collisions, characterized by a large momentum transfer, allow the simultaneous satisfaction of both the on-shell and short-time-scale conditions.
Quasi-free reactions with electromagnetic probes have been used to, for example, extract the neutron form factors \cite{Eden:1994ji,Bruins:1995ns,Ostrick:1999xa,E93026:2001css} and structure functions \cite{CLAS:2014jvt} from the deuteron. 
They have also been proposed for experiments at the planned electron-ion collider \cite{Jentsch:2021qdp,Friscic:2021oti}.
However, depending on the exact kinematic conditions, the contribution of multi-body knockout in experiments can be larger than naively expected \cite{Weinstein:1990rr}. In some instances, these final-state interactions are a source of background from the desired observables, while in others, they contain the signal of interest. In this work, we lay the foundation to understand final-state interactions in this general class of reactions in a model-independent fashion.  

A particularly interesting case is nuclear reactions with neutrons in the final state. Such reactions can be used to extract continuum properties of multi-neutron systems \cite{Duer:2022ehf}.
In Refs.~\cite{RIKENnnProp}, an alternative method to measure the neutron-neutron scattering length using the 
hard knockout reaction $^6$He$(p,p\alpha)nn$
in inverse kinematics was proposed. The method 
relies on the final-state interaction (FSI) of the 
neutrons after the sudden knockout of the $\alpha$ particle. The details of the neutron-neutron relative-energy distribution allow for a precise extraction of the scattering length if the kinematic conditions of the experiment are chosen appropriately. The expected uncertainty in the extraction of the scattering length was investigated by detailed calculations in Halo EFT \cite{Gobel:2021pvw}. 
Lazauskas et al. \cite{Lazauskas:2022mvq} used the sudden approximation to calculate the energy spectrum of the final state neutrons produced in the  hard knock-out reaction $^8$He$(p,p\alpha)4n$. Their calculation demonstrates that the resonance-like structure observed in the low-energy $4n$ spectrum \cite{Duer:2022ehf} can be generated by strong neutron-neutron dynamics in both the initial ${}^8$He bound and final $4n$ rescattering state. 

This scenario was extended to hard nuclear reactions with $N-1$ neutrons and one recoiled particle in the final state~\cite{Hammer:2021zxb}.
For interaction energies of the neutrons small compared to the energy scale of the primary reaction, factorization of the matrix element for the primary reaction and the final-state interaction of the neutrons was assumed. Moreover, it was assumed that the final-state interaction between the recoil particle and the neutrons is strongly suppressed because the recoiling particle has a large momentum relative to the $N-1$ neutrons. Using this set of assumptions, the differential cross section in the center-of-mass frame was written as 
\begin{equation}
\frac{d\sigma}{dE} \sim |M|^2\sqrt{E_{\rm NR}}\,\mathrm{Im}[G_{N-1}(E_\mathrm{kin}-E_\mathrm{NR},\mathbf{p})].
\label{eq:multi_n_factor}
\end{equation}
Here $M$ 
is a real function that is a production amplitude, $E_{\rm NR}$ 
is the non-relativistic energy of the recoil particle and $\mathbf{p}$ its momentum, while $E_{kin}$ is the total kinetic energy available to the final state particles and $\mathrm{Im}[G_{N-1}]$ is the imaginary part of the propagator of the multi-neutron state.
The above discussion assumes non-relativistic kinematics for all particles. However, extending the factorization assumption to relativistic kinematics is straightforward, as we will show below. For two neutrons, this treatment reduces to the Watson-Migdal treatment of final-state interactions \cite{Watson:1952ji,Migdal:1955}.

If $G_{N-1}$ is non-relativistic and
the energy of the recoil particle is close to its maximum possible energy, $\mathrm{Im}[G_{N-1}]$ is strongly constrained by Schr\"odinger/non-relativistic conformal symmetry \cite{Nishida:2007pj},
and the cross section follows a power law determined by the scaling dimension of the primary operator corresponding to the multi-neutron state \cite{Hammer:2021zxb}.

In this work, we leave the constraints from Schr\"odinger symmetry aside and instead focus on the proof of the factorization assumption and the suppression of final-state interactions with the recoil particle in such reactions, within a fully relativistic and model-independent formalism, while also providing its non-relativistic limit.
We consider four scenarios, all of which include an initial hadron struck by an external probe. The probe can either be a scalar current $(\Jc)$ or an additional hadron ($h_e$). In two of the scenarios, the initial hadron can be considered as a loosely bound state of two hadrons, and in the other two, it can be considered as a bound state of three hadrons. In both cases, the corresponding bound-state mass is denoted by $M_b$.  To distinguish these two cases, we will label the bound states as $H_{2}$ and $H_{3}$, respectively, where the subscript tells us the number of hadrons of which this can be considered a bound state of. More specifically, $H_{2}$ is a bound state of one light particle $\varphi$ and one heavy particle $\Phi$. Meanwhile, $H_{3}$ is a bound state of two $\varphi$ particles and one $\Phi$.  With these labels in place, we can enumerate the four scenarios we will consider:
\begin{itemize}
    \item \textbf{scenario 1}: $H_{2}+\Jc\to \varphi +\Phi$, given by the amplitudes labeled $\Tc_{1\to2}$,
    \item \textbf{scenario 2}: $H_{2}+h_e\to \varphi +\Phi +h_e$, given by the amplitudes labeled $\Tc_{2\to3}$,
    \item \textbf{scenario 3}: $H_{3}+\Jc\to 2\varphi +\Phi$, given by the amplitudes labeled $\Tc_{1\to3}$,
    \item \textbf{scenario 4}: $H_{3}+h_e\to 2\varphi +\Phi +h_e$, given by the amplitudes labeled $\Tc_{2\to4}$.
\end{itemize}
Note that the transition amplitudes are each exclusively labeled by the number of hadrons in the initial and final states, where the bound states are counted as a single hadron irrespective of their nature. Moreover, in the hadronic-probe examples we will consider $h_e=\varphi$.

In this work, we show that all of these transition amplitudes take on the following form:
\begin{align}
    \Tc = \Tc^{\mathrm{(LO)}} +\Tc^{\mathrm{(ps)}},
\end{align}
where $\Tc^{\mathrm{(LO)}}$ has a nearby pole associated with the nearly on-shell constituent hadrons and $\Tc^{\mathrm{(ps)}}$ is a less singular function, and consequently is relatively power suppressed. 

Furthermore, we derive closed-form expressions for $\Tc^{\mathrm{(LO)}} $, making a small set of approximations. First, we assume that the initial $H_i$ 
hadron can be represented as a loosely bound state of other hadronic constituents.  Second, we consider kinematics where the remnant $\varphi$ particles carry small momentum, compared to the struck $\Phi$ particle. Third, we assume that all particles involved are spinless and we use scalar currents. 

The first two assumptions allow us to arrive at a closed form for the amplitude in terms of the scattering amplitude of the remnant system while ignoring contributions that are kinematically suppressed. These corrections are generally more difficult to evaluate. The third assumption is for simplicity, and it is straightforward to lift.

The main results derived here all stem from the fact that, for kinematics where the struck particle is nearly on-shell before the collision, that particle's on-shell pole gives the dominant contribution to the amplitude.  
For the first two scenarios considered, there is only one particle left behind, in which case the residue of this pole can be written exactly in terms of the vertex function for the initial hadronic state to decay to its constituents and the on-shell form factor (scattering amplitude) if the probe is a current (hadron-hadron interaction). If more than one particle is left behind, then the on-shell scattering amplitude of the remnant $N-1$-body system also enters the residue. Away from the kinematics where the struck particle is nearly on shell, one can no longer assume that this pole contribution dominates the amplitude.

With this in mind, in Sec.~\ref{sec:main}, we begin with a brief discussion of the kinematic region of validity and then present our main result. In Sec.~\ref{sec:T_LO}, we present the derivation of the main results for scenarios 1, 2, 3, and 4. In all scenarios, we provide a fully relativistic derivation using all-order perturbation theory, as well as accompanying non-relativistic quantum mechanics. These complementary derivations provide an explicit dictionary between the amplitudes appearing in the relativistic framework and the wavefunctions and matrix elements appearing in the non-relativistic approach. In Sec.~\ref{sec:subleading}, we explain using arguments from Landau singularities~\cite{Landau:1959fi} why the remaining contributions to the amplitudes are less singular than the isolated pole contribution, which is sufficient to prove that these are kinematically suppressed relative to the contributions highlighted in the main results. In Secs.~\ref{sec:NR_loops_2B} and \ref{sec:NR_loops_3B}, we provide a detailed discussion of suppressed contributions using non-relativistic quantum mechanics. In Appendix~\ref{app:EFT_example}, we perform an explicit calculation of the one-neutron breakup in a halo nucleus using the non-relativistic effective Lagrangian presented in Ref.~\cite{Hammer:2017tjm}. In this explicit calculation, we reproduce the leading-order result obtained using the relativistic framework, but also obtain an explicit determination of the next-to-leading corrections. We find that these are indeed kinematically suppressed as claimed using the Landau singularity analysis.

\section{Main Results}
\label{sec:main}

In this section, we present our main results and the kinematics for which they hold. The four scenarios outlined in the introduction can be bundled into two sets of two. The first two scenarios include one particle left behind, with the distinction being the nature of the probe. We can generally label these two amplitudes as $\Tc_{n\to 1+n}$, where $n=1$ corresponds to the scalar current case and $n=2$ corresponds to the hadronic probe case. Similarly, scenarios 3 and 4 can be generically written as $\Tc_{n\to 2+n}$, where the $2$ in the subscript emphasizes that two particles were left behind to interact among themselves. Here we give expressions for $\Tc_{n\to 2+n}^{\mathrm{(LO)}}$, leaving the actual derivation for Sec.~\ref{sec:T_LO}. In Sec.~\ref{sec:subleading}, we explain why $\Tc_{n\to 2+n}^{\mathrm{(ps)}}$ is kinematically suppressed.

\subsection{Nearly on-shell Kinematics}
\label{sec:kin}

\begin{figure}[t]
    \centering
    \includegraphics[width=.8\linewidth]{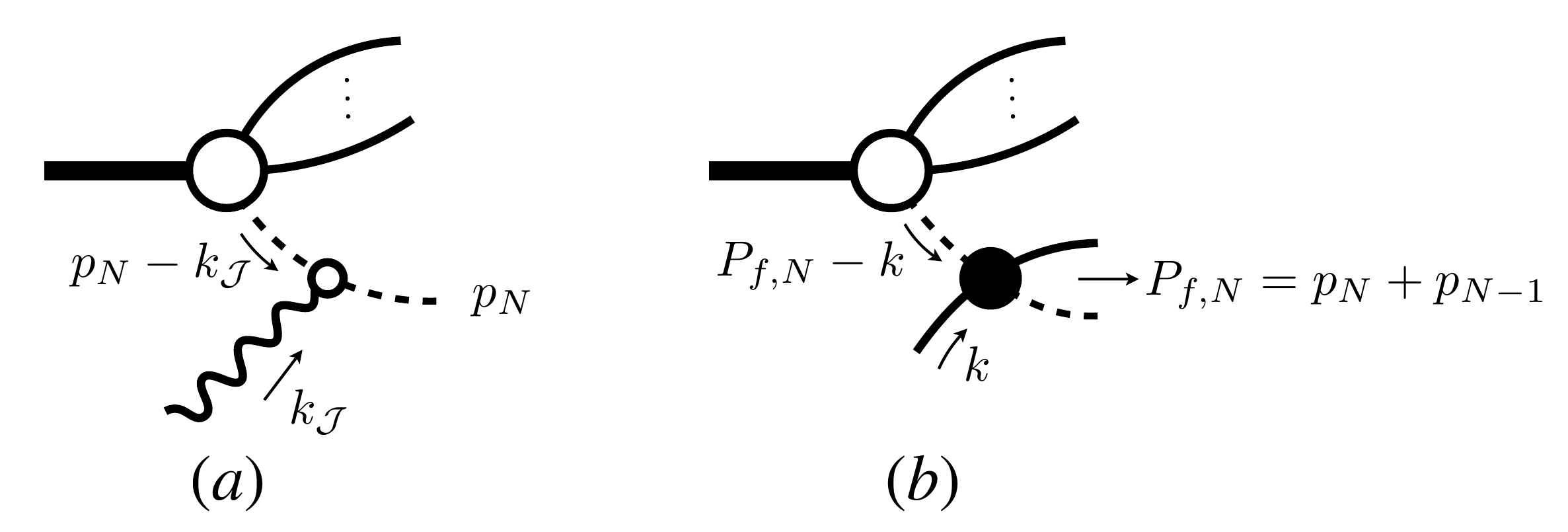}
    \caption{Leading-order diagrams contributing to the transition of the bound state to N particles. Diagram ($a$) illustrates a transition mediated by a current, while diagram (b) shows a transition mediated by a hadron. The thick solid line is the initial bound state, dashed lines represent the heavy particle in the final state, $\Phi$, the solid lines are the light particles, $\varphi$, the ``squiggly" line denotes the scalar current, and the different vertices are defined in  Fig.~\ref{fig:bbs}.
    }\label{fig:kinematics}
\end{figure}

We begin by reviewing the kinematics of interest. The generic class of amplitudes we are considering is qualitatively shown in Fig.~\ref{fig:kinematics}. In these two figures, we see an incoming particle, represented as a bold solid line to the left, that goes through a virtual decay to $N$ particles. This is not a real decay, because the incoming hadron is stable. In both figures, we see that one of the particles is struck by the probe. In Fig.~\ref{fig:kinematics}($a$), the probe appears as a wavy line, representing the scalar current. In Fig.~\ref{fig:kinematics}($b$), the probe is another hadron, depicted as another solid line. In all of the scenarios, we will consider the case where the leading effect arises from the probe striking the heavy particle $\Phi$ (e.g., the core of the nucleus) with mass $M$. The rest of the $\varphi$ particles will be assumed to be identical with a mass $m$. 

Given the important difference between these probes, they require different kinematic cuts. Let us look at each case individually first, before we proceed to bundle them in the rest of the analysis.

\subsubsection{External current scenario}

First, let us consider the scalar current case, since this is the simplest. Although in what follows we will be arriving at fully relativistic expressions, we are interested in low-energy systems. Furthermore, we will consider an external probe carrying spacelike four-momentum $k_{\Jc} = (\omega_\Jc, {\bf k}_{\Jc})$, where ${\bf k}_{\Jc} \gg \omega_\Jc \approx \frac{{\bf k}_{\Jc}^2}{2M}$. For these conditions to hold, the probe momentum has to satisfy $|{\bf k}_{\Jc}|\ll 2M$.

Now let $p_N=(\sqrt{M^2 +{\bf p}^2_N},{\bf p}_N)$ be the four-momentum of the final struck on-shell particle. It is easy to see that
\begin{align}
    (k_{\Jc}-p_N)^2-M^2
    &=k_{\Jc}^2+ p^2_N - 2k_{\Jc}\cdot p_N-M^2 
\nn\\
 &= k_{\Jc}^2 - 2 k_{\Jc}\cdot p_N
\nn\\
 &= k_{\Jc}^2 - 2 \omega_\Jc \sqrt{M^2 +{\bf p}^2_N} + 2 \,{\bf k}_{\Jc}
  \cdot \,{\bf p}_N. 
  \label{eq:pole_rel}
  \end{align}
Using the relations between $\omega_\Jc$ and ${\bf k}_{\Jc}$ introduced above and further assuming that the particle in the final state is moving non-relativistically, $M^2 \gg {\bf p}^2_N$, one finds
\begin{align}
    (k_{\Jc}-p_N)^2-M^2
 &\approx 
 -
   {\bf k}_{\Jc}^2
 -
   {\bf k}_{\Jc}^2
  +2
  \,{\bf k}_{\Jc}
  \cdot \,{\bf p}_N
  \nn\\
 &= - 2\,{\bf k}_{\Jc} \cdot \left( {\bf k}_{\Jc} - \,{\bf p}_N\right).
  \label{eq:pole_NR}
\end{align}
From this, it is clear that when the final on-shell particle is collinear with the incoming probe, the exchange propagator shown in Fig.~\ref{fig:kinematics} diverges as $\left| {\bf k}_{\Jc} -  {\bf p}_N\right|^{-1}$. In principle, nothing prevents one from using Eq.~\eqref{eq:pole_rel} and including relativistic corrections here, even though such corrections are expected to be negligible for the systems of interest.

\subsubsection{External hadron probe}
We now turn to the scenario where the probe is an on-shell particle of $\varphi$ type, with mass $m$ and four momentum $k=(\omega_{m,k},{\bf k})=(\sqrt{{\bf k}^2+m^2},{\bf k})$. From Fig.~\ref{fig:kinematics}($b$), one sees that the exchange propagator has a denominator of the form $(P_{f,N}-k)^2-M^2$, where $P_{f,N}=p_N+p_{N-1}$ is the four-momentum of the final two-particle state. Note, the $N-1$ particle will be the on-shell $\Phi$, while the $N$ particle will be the fast recoiled $\varphi$ particle that scattered from the $\Phi$.
We use the fact that $(P_{f,N}-k)^2$ is Lorentz invariant to write this in the center of momentum (CM) frame of the final two-particle system. We denote kinematic variables in this frame with a $\star$. 

With this, we write the pole of the propagator as  
\begin{align}
    (P_{f,N}-k)^2-M^2
    &= (P_{f,N}^\star-k^\star)^2-M^2
 \nn\\
    &= (P_{f,N}^\star)^2
    +(k^\star)^2- 2k^\star\cdot P^\star_f-M^2
 \nn\\
    &= s_{f,N}+m^2- 2\sqrt{s_{f,N}} \sqrt{(k^\star)^2+m^2}-M^2
 \nn\\
    &= s_{f,N} + m^2- 2\sqrt{s_{f,N}}\, \omega_{m,k}^\star -M^2
 \nn\\
    &= \sqrt{s_{f,N}}\left(
   \sqrt{s_{f,N}}
   - 2 \omega_{m,k}^\star
   +
   \frac{m^2-M^2}{\sqrt{s_{f,N}}}
\right)
\label{eq:expole},
\end{align}
where $\omega_{m,k}^\star =\sqrt{(k^\star)^2+m^2}$ and $s_{f,N}=P_{f,N}^2=E^{\star2}_f$ is the CM energy of the two-particle system.

Note, in the on-shell limit, $k^\star\to q^\star$, where $q^\star$ is defined by 
$\sqrt{s_{f,N}} = \sqrt{(q^\star)^2+m^2}
+\sqrt{(q^\star)^2+M^2}$, and one can see that Eq.~\eqref{eq:expole} vanishes. As a result, Fig.~\ref{fig:kinematics} will diverge as $|k^\star-q^\star|^{-1}$. Phenomenologically, one can enhance this on-shell condition by minimizing $|s_{f,N}- s_{\rm on}|$, where $ s_{\rm on}$ is defined as
\begin{equation}
 s_{\rm on} =    
\left(\omega_{m,k}^\star+\omega_{M,k}^\star\right)^2,
\end{equation}
where $\omega_{M,k}^\star =\sqrt{(k^\star)^2+M^2}$.
This is the would-be on-shell CM squared energy for the two-particle system composed of the external hadron probe and the nearly on-shell intermediate hadron.

In all the scenarios considered here, it will be important to further restrict the momenta of the $\varphi$-type particles. Let $p_j$ be the momentum of the $jth$ $\varphi$ particle, where $j<N$. Then the condition that all momenta must satisfy can be summarized as
\begin{align}
    |{\bf k}_{\Jc} | \sim |{\bf k} | \sim |{\bf p}_{N} | \gg |{\bf p}_{j} |.
    \label{eq:Momenta_hiarchy}
\end{align}
This hierarchy is understood in the rest frame of the initial bound state, where the probe injects large momentum and the spectator particles remain soft.

\subsection{Main result for $\Tc_{n\to 1+n}^{\rm (LO)}$}

The closed form for the amplitude strongly depends on the number of particles in the final state. In what follows, the most important distinguishing factor is the number of hadrons making up the initial bound state. First, we consider the cases where the initial hadron, $H_{2}$, can be thought of as a loosely bound state of two hadrons $(\varphi,\Phi)$ with masses $(m,M)$. For both external probes, these two hadrons will have momenta $(p_1,p_2)$. As already explained above, the two relevant transition amplitudes can be written as $\Tc_{n\to 1+n}$. 

We first consider the case where the incoming hadron is struck by a scalar current. In this case, the leading-order contribution to the transition amplitude can be written as, 
\begin{align}
i\Tc_{1\to 2}^{\rm (LO)}
=i\Gamma_2(k_{\Jc,2}^2)\bigg|_{k_{\Jc,2}^2 = M^2} \, \frac{i}{k_{\Jc,2}^2-M^2}
\,i f(k_\Jc^2),
\label{eq:case1}
\end{align}
where $k_{\Jc,2} = p_2-k_\Jc$, $p_2$ is the momentum of the struck particle, and 
 $\Gamma_2(M^2)$ is the purely hadronic coupling of the initial hadron to the final two hadrons, which can be taken to be on their mass shell. The pole comes from the propagator, and $f(k_\Jc^2)$ is the scalar form factor of the nearly on-shell particle.  It is important to note that the definition of the form factor, $f(k_\Jc^2)$, is normalized to $2M$ times the scalar charge~\footnote{The factor of $2M$ is to ensure that the charge is dimensionless, assuming that the current is given by the mass term in the Lagrangian of the theory.}, as opposed to $1$. An exact definition is given in Sec.~\ref{sec:LO_2Body}. Because we are interested in kinematics where the propagator is nearly on shell, both the $\Gamma_2$ function and the form factor $f$ can be safely evaluated on-shell. The error associated with this approximation can be absorbed in $\Tc_{1\to 2}^{\rm (ps)}$, where ``ps” denotes power-suppressed contributions.  

For the case where the probe is an on-shell hadron, we already mentioned that its incoming momentum is labeled ${\bf k}$. Let its outgoing momentum be ${\bf p}_3$. As a result, we can obtain the scattered angle ($\theta$) using the standard relation, ${\bf p}_3\cdot {\bf k} = |{\bf p}_3|\,|{\bf k}|\,\cos \theta$. With this, we can write the expression for $\Tc_{2\to 3}^{\rm (LO)}$ as
\begin{align}
i\Tc_{2\to 3}^{\rm (LO)}
=i\Gamma_2(k_{23}^2)\bigg|_{k_{23}^2 = M^2}\, \frac{i}{k_{23}^2-M^2}
\,i \Mch(P_{23}^2,\theta_{23}),
\label{eq:case2}
\end{align}
where $P_{23}=p_2+p_3$, $k_{23}=P_{23}-k$, the subscript $h$ in the scattering amplitude emphasizes that this is the scattering amplitude involving the \emph{hard} process between particles $2$ and $3$ of the final state. This will distinguish it from $\Mcs$, which will appear below to denote the scattering amplitude of \emph{slow}-moving particles. Again, this amplitude is on shell, with any off-shell dependence being absorbed into the definition of $\Tc_{2\to 3}^{\rm (ps)}$. 

\subsection{Main result for $\Tc_{n\to 2+n}^{\rm (LO)}$}

Next, we consider the case where $H_{3}$ can be thought of as a loosely bound state of three hadrons $(\varphi,\varphi, \Phi)$ with masses $(m,m,M)$. By restricting ourselves to the nearly on-shell kinematics discussed in Sec.~\ref{sec:kin}, $\Tc_{n\to 2+n}^{\rm (LO)}$ can be written as
\begin{align}
\Tc_{n\to 2+n}^{\rm (LO)}
&=
\boldsymbol{\Tc}^{(\mathrm{LO})}_{n\to 2+n, \mathrm{PV}}\bigg(1+\rho \,i\Mcs(s_{12})\bigg)
\nn\\
&\equiv
\Ac_{n\to 2+n}\,\Mcs(s_{12}),
\label{eq:3B_main_v0}
\end{align}
where $\Mcs$ is the two-particle scattering amplitude of the $\varphi+\varphi$ system which has total momentum $P_{12}=p_1+p_2$ and $s_{12}=P_{12}^2$. The subscript $s$ emphasizes that the particles that are scattering are the \emph{slow}-moving particles left behind. Above, $\rho$ is the two-body phase space of the slow-moving particles. An explicit definition of this is given in Sec.~\ref{sec:LO_3Body}.  

In Eq.~\eqref{eq:3B_main_v0}, we have chosen to present the result in two identical forms, which provide complementary intuition into the expression. The first of these expressions, we write in terms of a quantity labeled $\boldsymbol{\Tc}^{(\mathrm{LO})}_{n\to 2+n, \mathrm{PV}}$. This is defined diagrammatically in Sec.~\ref{sec:T_LO} to include a subclass of diagrams contributing to the full amplitude that have been evaluated using the principal value prescription, hence the subscript $\mathrm{PV}$. Using this function, one sees that the multiplicative factor includes the identity $1$ and a term proportional to the two-body phase space, $\rho$, and the two-body scattering amplitude, $\Mcs$. This expression suggests that in the limit that $\Mcs$ is perturbatively small, the identity term will dominate. In the second equality, we simplify this expression by introducing a new function $\Ac_{n\to 2+n}$, whose definition is explicitly given in Sec.~\ref{sec:T_LO}, but which can be reconstructed from this equality. It is important to emphasize that these two expressions are exactly identical, and it is just a matter of taste which one is used in the analysis. Furthermore, both of these functions are generally scheme-dependent quantities, and as a result, one must be quite careful when matching to other possible schemes. This can be done by matching the expressions of the on-shell scattering amplitudes, which are, by definition, scheme independent. Given the slightly simpler expression when using $\Ac_{n\to 2+n}$, we will discuss this one, even though the quantity that most naturally emerges from a derivation is $\boldsymbol{\Tc}^{(\mathrm{LO})}_{n\to 2+n, \mathrm{PV}}$.

The exact expression for the production amplitude, $\Ac_{n\to 2+n}$, depends on whether the probe is a current or a hadron, but for both cases, this includes the previously mentioned pole. The remainder, $\Tc_{n\to 2+n}^{\rm (ps)}$ is the ``\emph{power-suppressed}" contribution. We dedicate Sec.~\ref{sec:subleading} to explain why $\Tc_{n\to 2+n}^{\rm (ps)}$ is less singular and consequently is relatively suppressed. 

For the case where the external probe is a current, the production amplitude can be written as 
\begin{align}
i\Ac^{\Jc}_{1\rightarrow 3}=i\Gamma_3(P^2_{12},k^2_{\Jc,3})\bigg|_{k_{\Jc,3}^2=M^2}\, \frac{i}{k_{\Jc,3}^2-M^2}
\,i f(k_\Jc^2),
\label{eq:case3}
\end{align}
where 
$k_{\Jc,3} = p_3-k_{\Jc}$ is the inferred momentum of particle 3 before it is struck by the probe. 
The structure of this can be read off from Fig.~\ref{fig:kinematics}($a$). The $\Gamma_3$ function parametrizes the energy-dependent coupling of the initial hadron to the final three hadrons. The pole comes from the propagator, and $f(k_\Jc^2)$ is the scalar form factor of the nearly on-shell particle. Again, because we are interested in kinematics where the propagator is nearly on shell, both the $\Gamma_3$ function and the form factor $f$ can be safely evaluated on-shell. The error associated with this approximation can be absorbed in $\Tc_{1\to 3}^{\rm (ps)}$.

When the probe is changed for a hadron, the number of final hadrons is increased by one. We will label the momentum of the fourth hadron as $p_4$, which will have a mass $m$. For this case, the production amplitude can be written as, 
\begin{align}
i\Ac^{h}_{2\rightarrow 4}=i\Gamma_3(P_{12}^2,k_{34}^2)\bigg|_{k_{34}^2=M^2}\, \frac{i}{k_{34}^2-M^2}
\,i \Mch(P_{34}^2,\theta_{34}),
\label{eq:case4}
\end{align}
where $k_{34} = P_{34}-k$, $P_{34}=p_3+p_4$ and $\theta_{34}$ is the scattering angle, which is defined by ${\bf p}_4\cdot {\bf k} = |{\bf p}_4|\,|{\bf k}|\,\cos \theta_{34}$. 

In Eq.~\eqref{eq:3B_main_v0}, we have left the angular dependence of $\Mcs$ implicit. At low $s_{12}$, it is most convenient to partial-wave project it, since it will be dominated by the lowest partial waves. Making the angular momentum index $\ell$ explicit, we can write Eq.~\eqref{eq:3B_main_v0},
\begin{align}
\Tc_{n\to 2+n}^{\rm (LO)}
=
\sum_{\ell}
\Ac_{n\to 2+n}^{(\ell)}\,\Mcs^{(\ell)}(s_{12})
,
\label{eq:case4v2}
\end{align}
where we have used the fact that these amplitudes are independent of the magnetic quantum number $m$.

\section{Leading-order contributions}
\label{sec:T_LO}

In this section, we derive the leading-order contributions to the amplitudes. Given that our attention is focused on the kinematics discussed in Sec.~\ref{sec:kin}, this amounts to isolating the pole contribution to the amplitude. For scenarios 1 and 2, where there is only one particle left behind, this might seem trivial. That said, it is worth working through this exercise to clarify the procedure for placing non-perturbative functions on shell. We begin with scenario 1.

\begin{figure}[t]
    \centering
    \includegraphics[width=.9\linewidth]{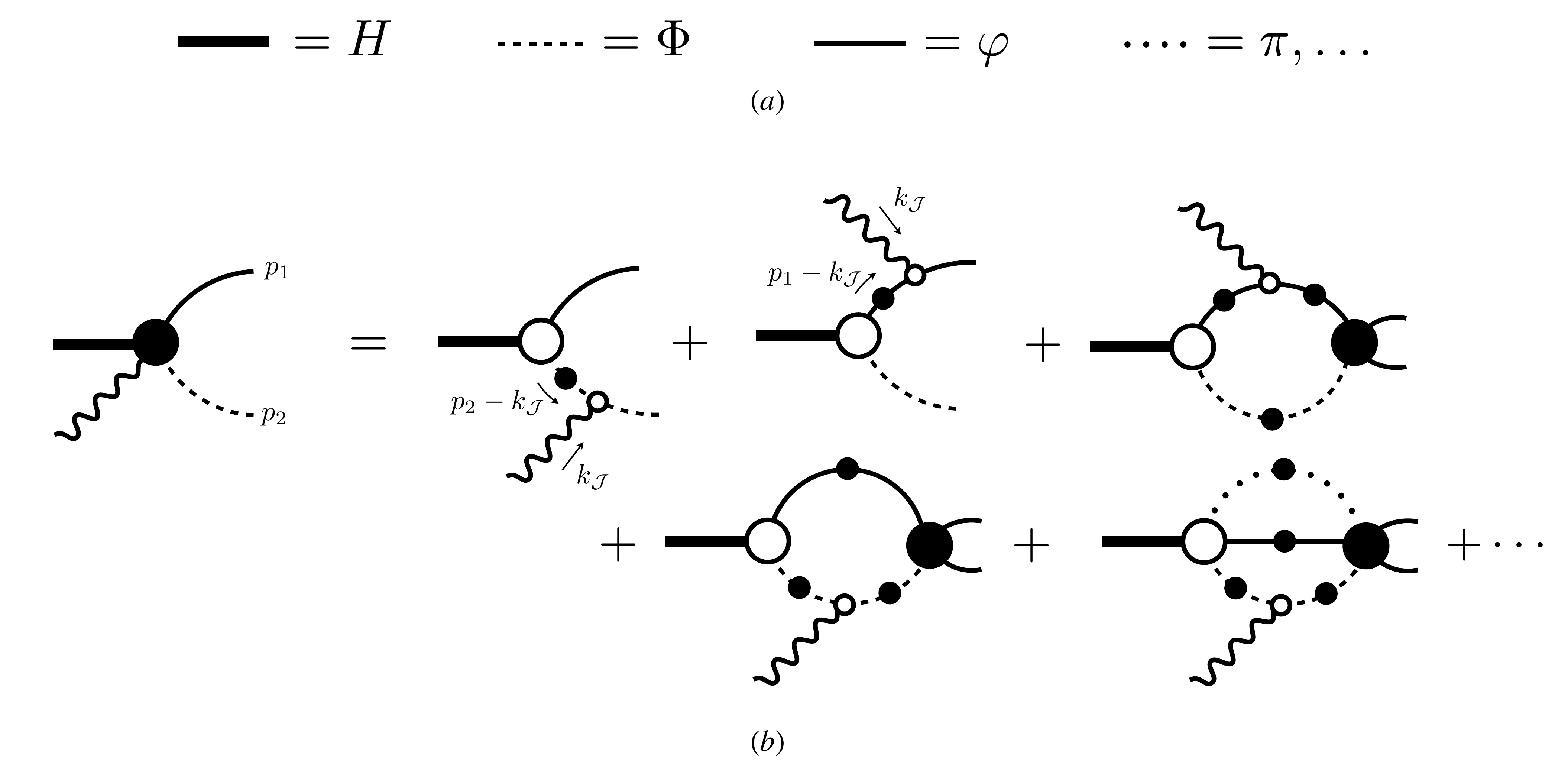}
    \caption{($a$) Shown is a reminder of the symbols for each particle type, previously introduced in Fig.~\ref{fig:kinematics}. ($b$) Shown is the diagrammatic representation of $i\Tc_{1\to 2}$ in terms of fully dressed vertices and propagators, which are themselves defined in Fig.~\ref{fig:bbs}. This amplitude includes the sum over all $1\to 2$ diagrams mediated by the insertion of a current, with all propagators and vertices being dressed. Shown here are just a subset of such diagrams. The leading-order contribution, and the focus of this work, is the first diagram to the right of the equality. Although diagrams that include effective two-body contributions to the current are not explicitly shown, their contributions are discussed in Sec.~\ref{sec:subleading}.}
    \label{fig:iT12}
\end{figure}

\subsection{Proof of Eq.~\eqref{eq:case1} for $\Tc_{1\to 2}^{\rm (LO)}$}
\label{sec:LO_2Body}

Before discussing the diagrammatic representation of $\Tc_{1\to2}$, we provide the matrix element definition of this amplitude using the standard relativistic definition of single-particle states. We label these states as $|H_2;{\bf P}\rangle_{\rm rel.}$, where the first label denotes the particle type and the second is the spatial momentum. Because the particles are on shell, their energy is fixed by their mass and momentum. Giving $H_2$ a mass $M_b$, its energy is $\omega_{M_b,P}=\sqrt{{\bf P}^2+M_b^2}$, and these states are normalized as,
\begin{equation}
    \langle H_2;{\bf P}'|H_2;{\bf P} \rangle_{\rm rel.}=2 \omega_{M_b,P}\,(2 \pi)^3 \delta^{(3)}({\bf P}'-{\bf P}),
    \label{eq:rel_norm}
\end{equation}
With this, we define the full $\Tc_{1\to2}$ amplitude as,
\begin{align}
\Tc_{1\to2}(p_1,p_2;P_i)
\equiv
\langle {\bf p}_1,{\bf p}_2;{\rm out}|
\Jc(0)
    |H_2;{\bf P}_i \rangle_{\rm rel.},
    \label{eq:rel_mat_elem}
\end{align}
where we have introduced the asymptotic out-state for a $\varphi+\Phi$ pair with momenta ${\bf p}_1$ and ${\bf p}_2$, and we have placed the current at the spatial origin. The reason for the latter is to remove the standard factor of $(2\pi)^4\delta^{(4)}(P_i +k_{\Jc}-P_f)$, where $P_f=p_1+p_2$, which explicitly imposes momentum conservation. This means that $\Tc_{1\to 2}$ is defined as a relativistic scattering amplitude, as opposed to an element of the relativistic $T$-matrix.

We now turn to the diagrammatic representation of this amplitude. In Fig.~\ref{fig:iT12}, we show the key classes of diagrams contributing to this process. Before focusing on the first diagram to the right of the equality, which contains the pole contribution we are after, we explain the philosophy behind Fig.~\ref{fig:iT12}.

In general, $\Tc_{1\to2}$ includes all diagrams coupling the initial and final states via a single current insertion. In drawing these diagrams, we treat the final-state hadrons as fundamental degrees of freedom, while the initial-state hadron is not itself a fundamental degree of freedom. Instead, it arises as a dynamical pole generated by the interactions among its constituents, encoded here as a pole in the two-particle scattering amplitude between $\varphi$ and $\Phi$. This is why there are no explicit diagrams associated with an $H_i$ pole in the final state; such poles are instead embedded in the final-state interactions, which are discussed further in Sec.~\ref{sec:subleading}.

\begin{figure}[t]
    \centering
    \includegraphics[width=.8\linewidth]{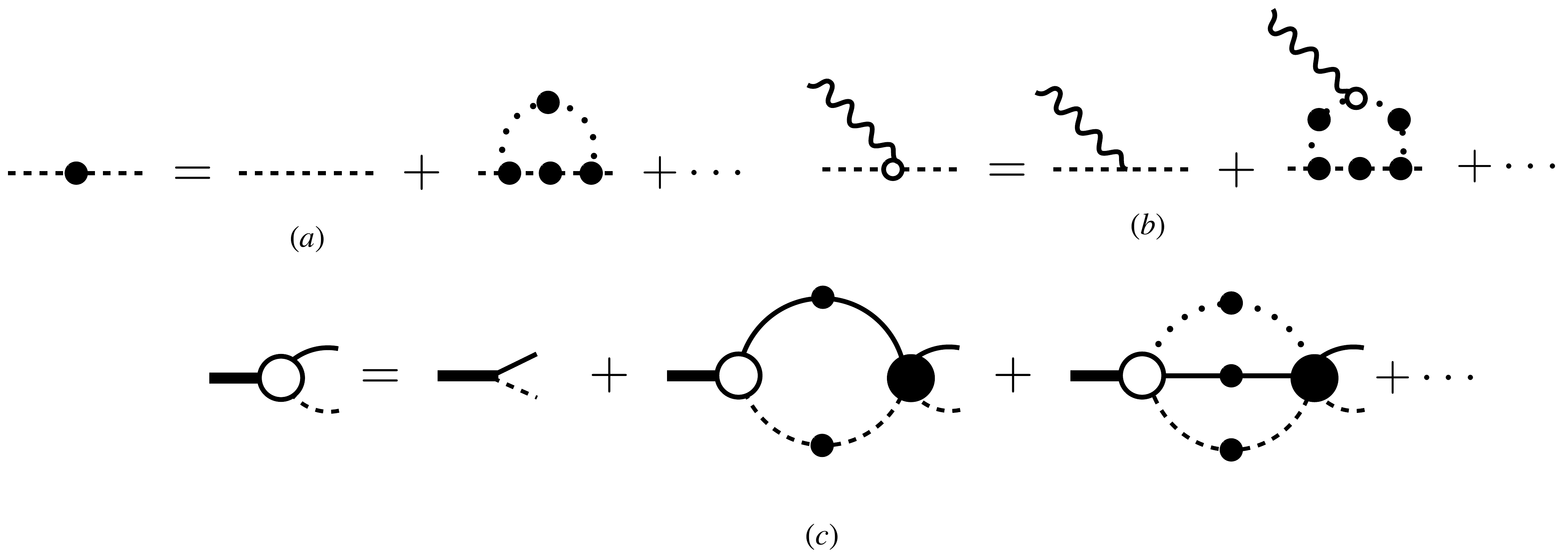}
    \caption{($a$) The fully dressed propagator includes the sum over all self-energy diagrams. Shown are just a subset of such diagrams. ($b$) Shown is the fully dressed off-shell form factor. ($c$) Shown is a self-consistent equation that the fully dressed $H\to \varphi+\Phi$ vertex must satisfy. Only a subset of possible diagrams are shown.}
    \label{fig:bbs}
\end{figure}

In drawing these diagrams, we keep in mind that all propagators and vertices are fully dressed, as made explicit diagrammatically. For example, the $\varphi$ and $\Phi$ propagators, depicted by solid and dashed lines respectively, carry filled circles to emphasize that all self-energy loops are included. Examples of self-energy contributions are shown in Fig.~\ref{fig:bbs}, where possible lighter intermediate particles are depicted as dotted lines.

Without loss of generality, the first diagram in Fig.~\ref{fig:iT12} can be written as,
\begin{align}
i\Tc_{1\to 2}^{(1)}
=i\Gamma_2(k_{\Jc,2}^2)\, i\Delta_\Phi\left(k_{\Jc,2}^2\right)
\,i f(k_{\Jc,2}^2,k_\Jc^2),
\label{eq:iT12_pole}
\end{align}
where $k_{\Jc,2}=p_2-k_\Jc$~\footnote{In general, we will set $k_{\Jc,n}=p_n-k_{\Jc}$. } is the momentum flowing through the $\Phi$ propagator, $\Delta_\Phi$ is the fully dressed propagator of $\Phi$, $f(k_{\Jc,2}^2,k_\Jc^2)$ is the off-shell form factor, and $\Gamma_2(k_{\Jc,2}^2)$ is a vertex function representing the partly off-shell $H_i \to \Phi + \varphi$ purely hadronic coupling. The fact that all particles and the current are scalars ensures there is only one form factor and only one vertex function. We now explain each of these in more detail.

The propagator of $\Phi$ can be written as,
\begin{align}
    i\Delta_\Phi\left(k_{\Jc,2}^2\right)
    =\frac{i}{k_{\Jc,2}^2-M^2 +i\epsilon}
    +iS(k_{\Jc,2}^2),
\end{align}
where $M$ is the physical mass, $S$ is a generally unknown function, and $\epsilon$ will be assumed to go to $0$ from above. We have used on-shell renormalization, so the residue equals $1$ at the physical pole. The function $S$ is non-singular in the vicinity of the pole; its singularities are associated with production thresholds of multi-particle states carrying the same quantum numbers as $\Phi$.

The off-shell form factor can be decomposed as,
\begin{align}
    f(k_{\Jc,2}^2,k_\Jc^2) 
     &= 
     f(M^2,k_\Jc^2)
     +
     \left[f(k_{\Jc,2}^2,k_\Jc^2)-f(M^2,k_\Jc^2)\right]
     \nn\\
     &\equiv 
     f(M^2,k_\Jc^2)
     +
     \delta f(k_{\Jc,2}^2,k_\Jc^2),
     \label{eq:off-shell-ff}
\end{align}
where $\delta f$ denotes the difference between the off-shell form factor and its on-shell value. The quantity $f(M^2,k_\Jc^2)$ is the on-shell physical elastic form factor of $\Phi$. Since $k_{\Jc,2}^2$ has been set to the constant $M^2$, we leave this argument implicit going forward and write,
\begin{align}
    f(k_{\Jc,2}^2,k_\Jc^2) 
     &=  f(k_\Jc^2)
     +
     \delta f(k_{\Jc,2}^2,k_\Jc^2).
     \label{eq:on-shell-ff}
\end{align}
The on-shell form factor $f(k_\Jc^2)$ is directly related to matrix elements of the current between on-shell single-particle states,
\begin{align}
    f({k}_\Jc^2) =  \langle \Phi;{\bf p}_2|\Jc(0)|\Phi;{\bf p}_2-{\bf k}_\Jc \rangle_{\rm rel.}.
    \label{eq:ff_def}
\end{align}
Similarly, the vertex function can be decomposed as,
\begin{align}
    \Gamma_2(k_{\Jc,2}^2) 
    = \Gamma_2 + \delta\Gamma_2(k_{\Jc,2}^2), 
\end{align}
where $\Gamma_2 \equiv \Gamma_2(k_{\Jc,2}^2)\big|_{k_{\Jc,2}^2=M^2}$ is the on-shell value of the vertex function. This is well-defined regardless of whether the purely hadronic process $H_i\to \Phi+\varphi$ is kinematically allowed. In the case of interest, it is not, since $H_i$ is a bound state with mass below the two-hadron threshold, $M_{b} < M + m$, and therefore cannot decay into two other hadrons.

Putting all these pieces together, we can rewrite Eq.~\eqref{eq:iT12_pole} as,
\begin{align}
i\Tc_{1\to 2}^{(1)}
&=i\Gamma_2 \frac{i}{k_{\Jc,2}^2-M^2} 
\,i f( k_\Jc^2) 
\nn\\
&\hspace{1cm}
-
\frac{i}{k_{\Jc,2}^2-M^2}\left(
\Gamma_2(k_{\Jc,2}^2)
\, f( k_{\Jc,2}^2 , k_\Jc^2)
- 
\Gamma_2\,
f( k_\Jc^2)
\right)
\nn\\
&\hspace{1cm}
- 
\Gamma_2(k_{\Jc,2}^2)
\, f( k_{\Jc,2}^2, k_\Jc^2)
\, i S ( k_{\Jc,2}^2)
\\
&\equiv i\Gamma_2 \frac{i}{k_{\Jc,2}^2-M^2} 
\,i f( k_\Jc^2) 
+
i\delta\Tc_{1\to 2}^{(1)},
\end{align}
where in the final line we have introduced $\delta\Tc_{1\to 2}^{(1)}$, whose definition follows from the prior lines. This quantity is finite in the vicinity of the $\Phi$ pole and therefore represents our first class of kinematically suppressed contributions, which can be ignored as one approaches the physical $\Phi$ pole. Note, in the kinematics being considered, it is safe to set $\epsilon$ explicitly in the definition of the propagator to $0$, since one never hits the pole. This establishes Eq.~\eqref{eq:case1} for scenario 1. The same procedure can be used to derive Eq.~\eqref{eq:case2} for scenario 2.
\subsection{Non-relativistic limit: two-body breakup}
\label{sec:2body_LO_NR}

We now turn to the non-relativistic limit of Eq.~(\ref{eq:case1}). We use non-relativistic quantum mechanics in a Hilbert space with a fixed number of particles to recompute the leading-order result established above in a quantum-field-theoretic formalism. This serves two purposes: first, it provides a dictionary to translate between relativistic and non-relativistic amplitudes; second, by connecting the result for quasi-free kinematics to existing literature, it allows us to use EFT arguments to estimate the power-suppressed corrections, as done in Secs.~\ref{sec:NR_loops_2B} and \ref{sec:NR_loops_3B} below.

We begin by defining the normalization of our non-relativistic one-body continuum state $|{\bf p}\rangle_{\rm NR}$ as
\begin{equation}
    \langle {\bf p}|{\bf p'} \rangle_{\rm NR}=(2 \pi)^3 \delta^{(3)}({\bf p}-{\bf p'}).
    \label{eq:NRnormn}
\end{equation}
The normalization of the state of $N$ non-interacting particles, $|\{{\bf p}'_i\}_{i=1}^N \rangle_{\rm NR}$,  directly follows from this definition. The one subtlety to bear in mind is that we always consider the case where $N-1$ of the particles are identical, while the $Nth$ particle is distinguishable from the rest. In what follows, we will need to define the $N$-particle eigenstate of the full Hamiltonian with outgoing boundary conditions, i.e., corresponding to $N$ particles with asymptotic momenta ${\bf p}_1, \ldots, {\bf p}_N$. We denote this state by $|\psi_{\{{\bf p}_i\}_{i=1}^N}^{(+)}\rangle_{\rm NR}$, and we impose that it is normalized identically to the free $N$ particle non-interacting eigenstate.

We introduce a non-relativistic analogue of $\Tc_{1\to 2}$, defined in Eq.~(\ref{eq:rel_mat_elem}), by placing the current at an arbitrary spacetime point $x$, integrating over this location to make momentum conservation explicit, and taking the non-relativistic limit of the initial and final states,
\begin{align}
   (2\pi)^4\delta^{(4)}(P_f-P_i-k_\Jc) \Tc_{1 \to 2}
   &=\int d^4 x\, e^{-ik_\Jc \cdot x} \,{}_{\rm rel.}\langle {\bf p}_1,{\bf p}_2;{\rm out}|
\Jc(x)
    |H_2;{\bf P}_i \rangle_{\rm rel.}
    \\
    &\approx
    \sqrt{2m}\sqrt{2M}\sqrt{2M_b}\int d^4 x\, e^{-ik_\Jc \cdot x} \, \langle \psi_{{\bf p}_1 {\bf p}_2}^{(-)}|
    \Jc(x)
    |{\bf P}_i \rangle_{\rm NR}
    \\
   & \equiv
    \sqrt{2m}\sqrt{2M}\sqrt{2M_b}\,\mathbf{T}_{1 \to 2}^{\rm NR},
    \label{eq:relNR}
\end{align}
where $\langle \psi_{{\bf p}_1 {\bf p}_2}^{(-)}|$ denotes the non-relativistic final-state bra corresponding to the relativistic out-state, and where we have introduced $\mathbf{T}_{1 \to 2}^{\rm NR}$. Identifying the current as a Heisenberg operator, $\Jc(x)=e^{i H t} \Jc(0,{\bf x})e^{-i H t}$, and performing the integral over $t$ analytically, this can be rewritten in terms of matrix elements of the Fourier-transformed current $\widetilde{\Jc}$,
\begin{align}
    \mathbf{T}_{1 \to 2}^{\rm NR}
    & = (2\pi)\,\delta( E_f -E_i - \omega_{\Jc})
    \int d^3 {\bf x}
    \, e^{+i {\bf k}_\Jc \cdot {\bf x}} \langle \psi_{{\bf p}_1 {\bf p}_2}^{(-)}|\Jc(0,{\bf{x}})|{\bf P}_i \rangle_{\rm NR}
    \label{eq:TNR_3d_int}
    \\
    & \equiv (2\pi)\,\delta(E_f - E_i-\omega_{\Jc})
    \langle \psi_{{\bf p}_1 {\bf p}_2}^{(-)}|\widetilde{\Jc}(0,{\bf k}_\Jc)|{\bf P}_i \rangle_{\rm NR}.
\label{eq:TNR1}
\end{align}

The translation property $\Jc(0,{\bf x})=e^{-i \hat{\bf{P}}\cdot \hat{\bf{x}}} \Jc(0)e^{+i \hat{\bf{P}}\cdot \hat{\bf{x}}}$ can be used to evaluate the remaining spatial integral in Eq.~\eqref{eq:TNR_3d_int}, yielding the remaining momentum-conserving delta function and reproducing Eq.~\eqref{eq:rel_mat_elem} in the non-relativistic basis. Instead, we proceed to address a subtlety specific to the non-relativistic setting.

In a relativistic QFT, particle number is not conserved, and currents can couple states in a Fock space with arbitrarily different numbers of particles. In contrast, in a non-relativistic theory, states with different particle numbers reside in different Hilbert spaces. It is therefore convenient to decompose the full current $\Jc$ into operators that each couple a specific pair of Hilbert spaces. Since the LO term of the amplitude is dominated by the single-particle pole, we can replace $\Jc$ with the one-body operator $\Jc_{\rm 1B}$ at leading order; the error is kinematically suppressed. This operator couples a single $\Phi$ state to another single $\Phi$ state.

\begin{figure}[t]
    \centering
    \includegraphics[width=.8\linewidth]{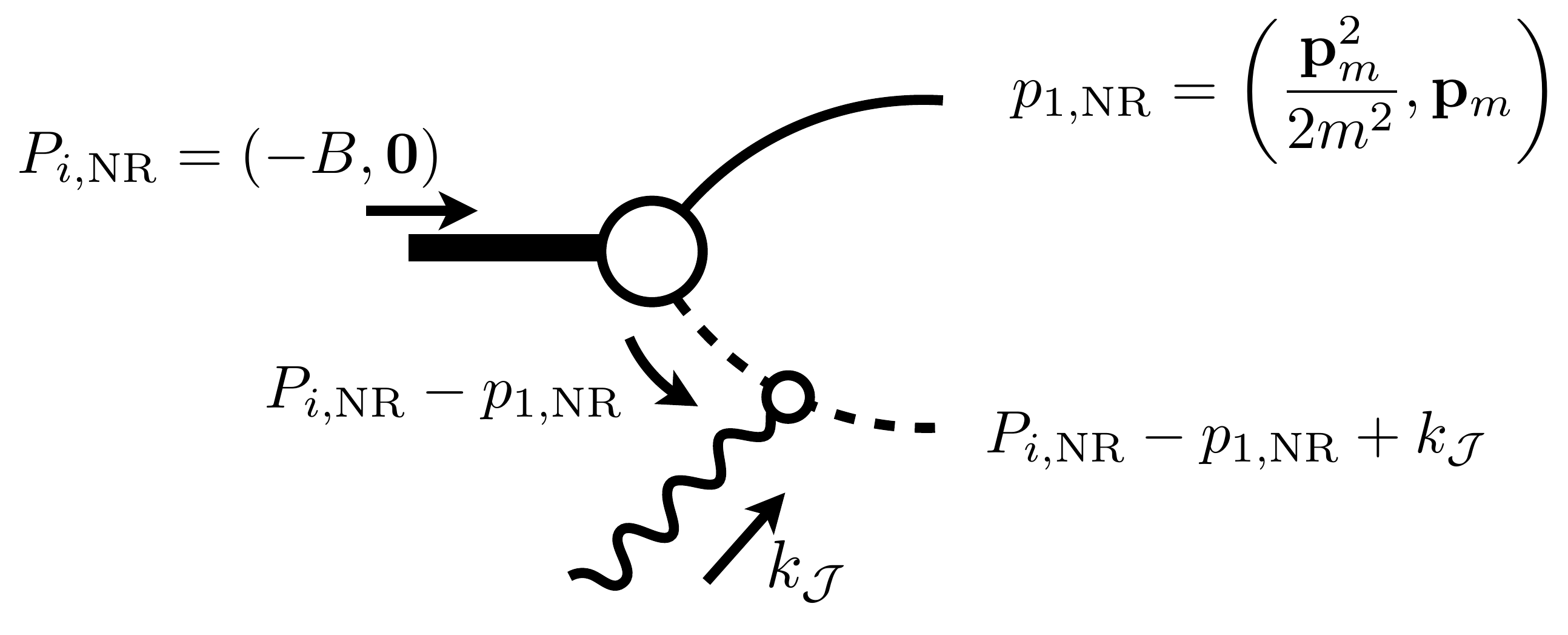}
    \caption{Shown is the tree-level dominant diagram using the non-relativistic kinematics defined in the text. The subscript $\rm NR$ emphasizes that the rest-mass energy has been removed from the four-momentum and only leading-order non-relativistic corrections are kept in the energy.}
    \label{fig:NR_1body_tree}
\end{figure}

In non-relativistic quantum mechanics, both $|{\bf P}_i \rangle_{\rm NR}$ and $|\psi_{{\bf p}_1 {\bf p}_2}^{(+)}\rangle$ can be factored into a center-of-mass part and an internal wave function, exploiting the fact that translation-invariant interactions commute with the total-momentum operator,
\begin{eqnarray}
    |{{\bf P}}_i \rangle_{\rm NR} &=&|{\bf P}_i \rangle_{\rm tot, NR} \otimes |\Psi_B^{{\rm int}} \rangle, \nonumber\\
    |\psi_{{\bf p}_1 {\bf p}_2}^{(+)}\rangle&=&|{\bf P}_f \rangle_{\rm tot, NR} \otimes|\psi_{{\bf p}_{\rm rel}}\rangle. \label{eq:internal}
\end{eqnarray}
The state $|\Psi_B^{\rm int} \rangle$ is a stationary eigenstate of the internal part of $\hat{H}$ (i.e., with center-of-mass motion removed) corresponding to energy $E^{\rm int}=-B$, where $B>0$ is the binding energy of the two-body bound state. It is normalized to one. The corresponding internal eigenstate in the final state has energy $E_f - {\bf P}^2_f/[2(m+M)]={\bf p}_{\rm rel}^2/(2 m_R)$, with $m_R =mM/(m+M)$ being the reduced mass. The energy-conserving delta function in Eq.~(\ref{eq:TNR1}) then requires that, in the rest frame of the bound state as depicted in Fig.~\ref{fig:NR_1body_tree},
\begin{equation}
    \omega_{\Jc}-B= E_f = \frac{{\bf P}^2_f}{2 (m+M)} + \frac{{\bf p}_{\rm rel}^2}{2 m_R}.
\label{eq:omega}
\end{equation}

We now factorize the current operator in the same spirit as Eq.~(\ref{eq:internal}). In the kinematics in which ${\bf k}_J$ is approximately equal to the final-state momentum of the $\Phi$ particle, the dominant contribution comes from the current coupling to particle 2 (the struck particle $\Phi$). We therefore write, 
\begin{equation}
    \widetilde{\Jc}(0,{\bf k}_\Jc)=\mathbb{I}_1 \otimes \widetilde{\Jc}^{\rm 1B}_{2}(0,{\bf k}_\Jc)  + \cdots,
    \label{eq:Jcfactorization}
\end{equation}
where the ellipsis contains both the one-body current acting on particle 1 (i.e., $\widetilde{\Jc}^{\rm 1B}_{1}(0,{\bf k}_\Jc)\otimes \mathbb{I}_2$) and higher-body operators, neither of which contribute to the LO pole dominating quasi-free scattering. The one-body current acting on particle 2 is defined by:
\begin{equation}
    \langle {\bf p}_2'|\Jc^{\rm 1B}_{2}(0,{\bf k}_{\Jc})|{\bf p}_2\rangle_{\rm NR}=(2 \pi)^3 \delta^{(3)}({\bf p}_2'-{\bf k}_{\Jc}-{\bf p}_2) \frac{f(k_{\Jc}^2)}{2M},
    \label{eq:J1B}
\end{equation}
where $f(k_{\Jc}^2)$ is the form factor previously defined in Eq.~\eqref{eq:ff_def}. The factor of $2M$ in the denominator reflects the normalization of the non-relativistic basis, Eq.~\eqref{eq:NRnormn}, while assuring that $f(k_{\Jc}^2)$ is defined as in the relativistic formalism, Eq.~\eqref{eq:ff_def}. Note, this makes no assumption regarding the normalization of the form factor itself.   

Inserting Eqs.~(\ref{eq:Jcfactorization}), (\ref{eq:J1B}), and (\ref{eq:internal}) into Eq.~(\ref{eq:TNR1}), we find,
\begin{equation}
\mathbf{T}_{1 \to 2}^{\rm NR}=(2 \pi)^4 \delta(E_f - E_i - \omega_{\Jc}) \delta^{(3)}({\bf P}_f - {\bf k}_\Jc - {\bf P}_i) \frac{f(k_\Jc^2)}{2M} \langle \psi_{{\bf p}_{\rm rel}}^{(-)}|\identity_{\rm 1}|\Psi_B^{\rm int} \rangle_{\rm NR} + \ldots,
\end{equation}
    where $\langle\psi_{{\bf p}_{\rm rel}}^{(-)}|$ is the outgoing two-body scattering bra state for the $\varphi \Phi$ pair, and the ellipsis denotes contributions in which the probe couples to the two-body state rather than to particle 2 alone, i.e., two-body currents, or to particle 1. 

Inserting this into Eq.~(\ref{eq:relNR}), we find,
\begin{align}
    \Tc_{1 \to 2} \approx \sqrt{2m} \sqrt{2M} \sqrt{2 M_b} \frac{f(k_\Jc^2)}{2M} \langle \psi_{{\bf p}_{\rm rel}}^{(-)}|\identity_{\rm 1}|\Psi_B^{\rm int} \rangle_{\rm NR} + \ldots,
    \label{eq:NRderivationfull2B}
\end{align}
where the $\approx$ reflects that we have not used the full relativistic normalization of the states.

If final-state interactions are treated as subleading, the interacting final state $|\psi_{{\bf p}_{\rm rel}}^{(+)}\rangle$ reduces to a plane wave, and we obtain,
\begin{equation}
   \Tc_{1 \to 2} \approx \sqrt{2M_b} \sqrt{2m}\sqrt{2M} \langle {\bf p}_m|\Psi_B^{\mathrm{int}} \rangle \frac{f(k_\Jc^2)}{2M} + \Tc_{1 \to 2}^{\rm ps},
   \label{eq:nrscenarioI}
\end{equation}
where particle 2, $\Phi$, has been identified as the struck particle. The only approximation in the first term is in the final-state normalization, so Eq.~(\ref{eq:nrscenarioI}) is correct up to $\mathcal{O}(p_m^2/M^2)$. The power-suppressed remainder $\Tc_{1 \to 2}^{\rm ps}$ accounts for final-state interactions, two-body currents, and graphs where the probe strikes particle 1.

In the leading term, the identity operator acting on particle 1, together with the plane wave in the final state, allows us to identify the momentum of the spectator particle as ${\bf p}_m$:
\begin{equation}
    {\bf p}_m={\bf p}_1-\frac{m}{M+m}{\bf P}_i={\bf p}_{\rm rel}+\frac{m}{M+m}\bf{k}_{\Jc}.
\end{equation}
In the center-of-mass frame of the initial bound state, this implies that particle 2 has momentum $-{\bf p}_m$ before the interaction with the probe, as depicted in Fig.~\ref{fig:NR_1body_tree}. By convention, ${\bf p}_m$ is called the \emph{missing momentum}.

Because we have converted from the non-relativistic amplitude to the relativistic invariant amplitude, this result can be used in the standard relativistic relationship between the invariant amplitude squared and the differential cross section. The non-relativistic amplitude involves only the relative degrees of freedom and is interpreted most directly in the rest frame of the initial bound state; amplitudes in other frames are obtained by Galilean boosts, though kinematic relations such as Eq.~(\ref{eq:omega}) are frame dependent.

\subsection{Proof of Eq.~\eqref{eq:3B_main_v0} for $\Tc_{1\to 3}^{\rm (LO)}$}
\label{sec:LO_3Body}

We now move on to scenario 3 and derive the leading-order contribution to $\Tc_{1\to3}$. In this case, the heavy particle $\Phi$ in the final state carries momentum $p_3$, while the two light particles of type $\varphi$ carry momenta $p_1$ and $p_2$. The key departure from the previous case is that, for the kinematics being considered, the two $\varphi$ particles can interact arbitrarily strongly. In other words, the kinematics do not suppress the scattering amplitude of this subsystem, which we labeled $\Mcs$ in Eq.~\eqref{eq:3B_main_v0}. As a result, these interactions must be treated to all orders.

\begin{figure}[t]
    \centering
    \includegraphics[width=1\linewidth]{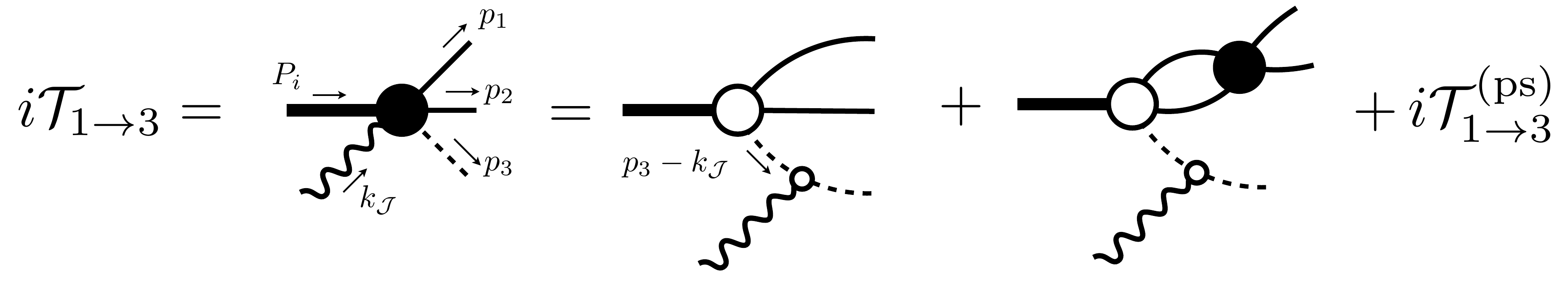}
    \caption{Diagrammatic representation of the amplitude $1+\mathcal{J}\rightarrow 3$ with a hard probe. In this and subsequent diagrams, the dressing of propagators is left implicit.}
    \label{fig:iT13_LO}
\end{figure}

The amplitude $\Tc_{1\rightarrow 3}$ is defined by the sum over all connected diagrams coupling the initial one-particle state to the final three-particle state via a single current insertion; see Fig.~\ref{fig:iT13_LO} for a diagrammatic presentation. We have simplified the diagrams in two ways compared to the previous section. First, the filled circles on the propagators are dropped, with dressing now understood to be implicit. Second, we only draw the leading-order contributions; subleading ones are absorbed into $\Tc^{\rm (ps)}_{1\to3}$, where the superscript indicates that these contributions are power-suppressed relative to those being kept.

We provide an all-orders-in-perturbation-theory proof that the dominant contribution satisfies a Watson-like theorem: its phase is given by the purely hadronic phase of the two-$\varphi$ subsystem, i.e., the phase of $\Mcs$. This holds when the center-of-mass energy of the two-$\varphi$ subsystem does not exceed the first inelastic threshold in that subsystem, $(p_1+p_2)^2<s_{\rm ine.}$. With this kinematic restriction, we can recover the Migdal-Watson treatment of final-state interactions~\cite{Watson:1952ji,Migdal:1955}. A recent, all-orders-in-perturbation-theory derivation of this result for two-particle systems in relativistic kinematics can be found in Ref.~\cite{Briceno:2020vgp}.

To arrive at the leading-order contribution, we introduce an intermediate object $\widetilde{\Tc}_{1\to3}^{(\mathrm{LO})}$. As will be shown below, $\widetilde{\Tc}_{1\to3}^{(\mathrm{LO})}$ includes the leading-order contribution we are after, as well as some residual suppressed corrections. We define $\widetilde{\Tc}_{1\to3}^{(\mathrm{LO})}$ to satisfy the following equation, see Fig.~\ref{fig:iT13_LO} for comparison,
\begin{align}
i\widetilde{\Tc}_{1\to3}^{(\mathrm{LO})} (P_i, k_\Jc; p_1, p_2)
    &= i \boldsymbol{\Tc}^{(0)}_{1\to3} (P_i, k_\Jc; p_1, p_2) + \xi \int_\ell\,i\boldsymbol{\Tc}^{(0)}_{1\to3}(P_i, k_\Jc; \ell,p_1+p_2-\ell)
    \nn\\
    &\times
    i\Delta_\varphi(\ell^2)\,
    i\Delta_\varphi(\ell_{12}^2)\, i\Mcsoff(\ell,p_1+p_2-\ell;p_1,p_2),
    \label{eq:Tau13}
\end{align}
where $\ell_{12} = p_1+p_2-\ell$, $\boldsymbol{\Tc}^{(0)}_{1\to3}$ is a real-valued function in the kinematic region of interest, and $\Mcsoff$ is the off-shell extension of the two-$\varphi$ scattering amplitude. We define
\begin{equation}
    \int_\ell \equiv \frac{{\rm d}^4 \ell}{(2\pi)^4},
\end{equation}
and introduce the shorthand for the product of two $\varphi$ propagators,
\begin{equation}
    \Delta^{(2\varphi)}(p;k) \equiv i\Delta_\varphi(p^2)\,i\Delta_\varphi(k^2).
\end{equation}
The quantity $\boldsymbol{\Tc}^{(0)}_{1\to3}$ can be written in terms of a real-valued function $\Kc_{1\to3,0}$, a fully-dressed propagator for $\Phi$, and the off-shell form factor $f$,
\begin{align}
i\boldsymbol{\Tc}^{(0)}_{1\to3}(P_i, k_\Jc; p_1, p_2)
=
i\Kc_{1\to3,0}(P_i, k_\Jc; p_1, p_2)\,i\Delta_\Phi(k_{\Jc,3}^2) \, i f(k_\Jc^2),
\label{eq:T130}
\end{align}
where $k_{\Jc,3}= p_3-k_\Jc $. The function $\Kc_{1\to3,0}$ is defined as the sum over all $1\to 3$ diagrams with no current insertion with a subset of diagrams missing. The class of diagrams missing are pair-wise $\varphi\varphi$ rescattering in the final state - the kind explicitly shown in the second term of the equality of Fig.~\ref{fig:iT13_LO}. It is worth emphasizing that this quantity is quite unorthodox, since some classes of rescattering diagrams are indeed present in its definition (e.g., those associated with $\varphi \Phi$ pair-wise scattering), while some associated with $ \varphi\varphi$ are absent. The diagrams that are missing in the definition of $\Kc_{1\to3,0}$, are explicitly shown in Fig.~\ref{fig:iT13_LO}. The subscript ``$0$'' counts the number of final-state insertions of the $\varphi\varphi\to \varphi\varphi$ Bethe-Salpeter kernel, to be introduced shortly.~\footnote{There are in fact an infinite number of $\varphi\varphi\to \varphi\varphi$ Bethe-Salpeter kernels inside of the definition of $\Kc_{1\to3,0}$, associated with intermediate pair-wise scattering that appears in standard ladder equations, see Refs.~\cite{Hansen:2015zga,Briceno:2019muc,Jackura:2019bmu,Jackura:2022gib,Draper:2023xvu, Briceno:2024ehy, Feng:2024wyg,Sadasivan:2021emk,Mai:2017wdv,Mikhasenko:2019vhk} for a modern discussion in such. The subscript discussed above counts the number of kernels that are being shown explicitly.} 

\begin{figure}[t]
    \centering
    \includegraphics[width=1\linewidth]{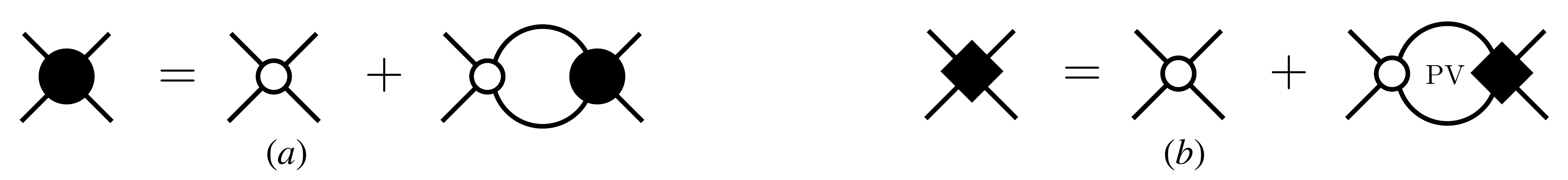}
    \caption{Diagrammatic representation of ($a$) the off-shell two-$\varphi$ scattering amplitude $i\Mcs$ and ($b$) the corresponding $K$-matrix $\Kcs$. }
    \label{fig:amps:kernel}
\end{figure}

The off-shell two-$\varphi$ scattering amplitude satisfies the integral equation,
\begin{equation}
    i\Mcsoff(p,k) = i\mathcal{B}(p,k) + \xi\int_{\ell} i\mathcal{B}(p,\ell)\,\Delta^{(2\varphi)}(\ell;P-\ell)\, i \Mcsoff (\ell, k),
    \label{eq:M2}
\end{equation}
depicted diagrammatically in Fig.~\ref{fig:amps:kernel}($a$). Here $\mathcal{B}$ is the renormalized $2\to2$ Bethe-Salpeter kernel, containing all two-particle-irreducible $s$-channel diagrams. It is a smooth function in the kinematic region where only two $\varphi$ particles can go on shell, a feature that will be important when isolating singularities in $\Tc_{1\to3}^{(\mathrm{LO})}$. The factor $\xi$ is the symmetry factor of the two-particle system: $\xi = 1/2$ for identical particles and $\xi = 1$ otherwise.

We now perform the on-shell decomposition of $\boldsymbol{\Tc}^{(0)}_{1\to3}$ in Eq.~\eqref{eq:T130}, following the same steps as in Sec.~\ref{sec:LO_2Body},
\begin{align}
i\boldsymbol{\Tc}^{(0)}_{1\to3}(P_i, k_\Jc; p_1, p_2)
&=
\left[i\Kc_{1\to3,0}(P_i, k_\Jc; p_1, p_2)\, i f(k_\Jc^2)\right]\bigg|_{k_{\Jc,3}^2 = M^2}
\frac{i}{k_{\Jc,3}^2-M^2}
+
\delta i\boldsymbol{\Tc}^{(0)}_{1\to3}(P_i, k_\Jc; p_1, p_2)
\nn\\
&\equiv
i\boldsymbol{\Tc}^{({\rm p})}_{1\to3}(P_i, k_\Jc; p_1, p_2)
+
\delta i\boldsymbol{\Tc}^{(0)}_{1\to3}(P_i, k_\Jc; p_1, p_2),
\label{eq:T130_v1}
\end{align}
where $i\boldsymbol{\Tc}^{({\rm p})}_{1\to3}$ denotes the pole piece of this production vertex. We now see that $\widetilde{\Tc}_{1\to3}^{(\mathrm{LO})}$ includes both the leading-order pole piece and additional off-shell corrections---as claimed above. We therefore define ${\Tc}_{1\to3}^{(\mathrm{LO})}$ to satisfy the same equation as $\widetilde{\Tc}_{1\to3}^{(\mathrm{LO})}$, but with the pole piece as the driving term,
\begin{align}
i{\Tc}_{1\to3}^{(\mathrm{LO})} (P_i, k_\Jc; p_1, p_2)
    &= i \boldsymbol{\Tc}^{(p)}_{1\to3} (P_i, k_\Jc; p_1, p_2) + \xi \int_\ell\,i\boldsymbol{\Tc}^{(p)}_{1\to3}(P_i, k_\Jc; \ell,p_1+p_2-\ell)
    \nn\\
    &\times
    \Delta^{(2\varphi)}(\ell;P-\ell)\, i\Mcsoff(\ell,p_1+p_2-\ell;p_1,p_2).
    \label{eq:Tau13p}
\end{align}
Note that $\boldsymbol{\Tc}^{(p)}_{1\to3}$ is not yet fully on shell, since it depends on the loop momenta of the two $\varphi$ particles being integrated.

We now evaluate the residue of the pole, which depends on $\Mcs$. Since the imaginary part of an amplitude is related to its singularities---which arise from intermediate particles going on-shell---deriving the phase of $\Tc_{1\to3}^{(\mathrm{LO})}$ amounts to isolating all of its singularities. For this, we use the identity~\cite{Briceno:2020vgp},
\begin{align}
    \xi\int_{\ell} i\Lc^{({\rm off})} (p,\ell)\,\Delta^{(2\varphi)}(\ell;P-\ell)\, i \Rc^{({\rm off})}  (\ell, k)
    &\nn\\
    &\hspace{-4cm}
    =i\Lc(p,q^\star)\,\rho \,i \Rc (q^\star, k)+
    \xi\int_{\ell,\rm PV} i\Lc^{({\rm off})} (p,\ell)\,\Delta^{(2\varphi)}(\ell;P-\ell)\, i \Rc^{({\rm off})}  (\ell, k),
    \label{eq:rho_PV}
\end{align}
where $\Lc^{({\rm off})}$ and $\Rc^{({\rm off})}$ are generic off-shell vertices coupling to two-particle states, $\rho=\xi q^\star/(8\pi\sqrt{s_{12}})$ is the two-body phase space, $q^\star = \sqrt{s_{12}/4-m^2}$ is the on-shell relative momentum, and PV denotes the principal-value prescription for the propagators. The term proportional to $\rho$ captures the sole kinematic singularity of this amplitude and is purely imaginary above threshold. Since this contribution arises from placing the two intermediate particles on shell, the vertices $\Lc$ and $\Rc$ multiplying it are on shell, and the superscript ``off'' can be dropped for that term. In angular-momentum space, where the phase space factor is proportional to the identity, this product reads $i\Lc(p,q^\star)\,\rho\,i\Rc(q^\star,k) = \sum_{\ell} i\Lc_{\ell }(p,q^\star)\,\rho\,i\Rc_{\ell }(q^\star,k)$.

Applying this identity to each $s$-channel loop in Eq.~(\ref{eq:M2}) and organizing the result by the number of $\rho$ insertions, we define the $K$-matrix as the quantity satisfying the same integral equation as $\Mcsoff$ but with all integrals replaced by the PV prescription,
\begin{equation}
    i\Kcsoff(p,k)  = i\mathcal{B}(p,k) + \xi\int_{\ell, \rm PV} i\mathcal{B}(p,\ell)\,\Delta^{(2\varphi)}(\ell;P-\ell)\, i \Kcsoff (\ell, k),
    \label{eq:K}
\end{equation}
depicted diagrammatically in Fig.~\ref{fig:amps:kernel}($b$). Placing the external legs on shell and summing over all $\rho$ insertions in Eq.~\eqref{eq:M2}, one finds,
\begin{align}
i\Mcs = i \Kcs \sum_{n=0}^{\infty}(\rho\,i\Kcs )^n
=
i \left[\Kcs^{-1} - i \rho\right]^{-1},
\label{eq:M2_K2}
\end{align}
where all quantities are understood as matrices in angular-momentum space. By rotational and azimuthal symmetry, $\Kcs$ and $\Mcs$ are diagonal in $\ell$ and independent of $m$. For further details, we refer the reader to Ref.~\cite{Briceno:2020vgp}.

We are now in a position to isolate all singularities of $\Tc_{1\to3}^{(\mathrm{LO})}$. Applying Eq.~\eqref{eq:rho_PV} to replace every internal loop in Eq.~\eqref{eq:Tau13p}, and collecting terms at zeroth order in $\rho$ insertions, we define,
\begin{align}
i\boldsymbol{\Tc}^{(\mathrm{LO})}_{1\rightarrow 3, \mathrm{PV}}
(P_i, k_\Jc; p_1, p_2)
    &\equiv i \boldsymbol{\Tc}^{(p)}_{1\to3} (P_i, k_\Jc; p_1, p_2) + \xi\int_{\ell,\rm PV}\,i\boldsymbol{\Tc}^{(p)}_{1\to3}(P_i, k_\Jc; \ell,p_1+p_2-\ell)
    \nn\\
    &\times \Delta^{(2\varphi)}(\ell;p_1+p_2-\ell)\, i\Kcsoff(\ell,p_1+p_2-\ell;p_1,p_2).
    \label{eq:Tau13LOPV_def}
\end{align}
Just like the $K$-matrix, this quantity is non-singular and purely real in this kinematic region, as assured by the PV prescription.

Summing over all insertions of $\rho$ and suppressing kinematic arguments (with all quantities understood as matrices in angular-momentum space), we obtain,
\begin{align}
    i\Tc_{1\to3}^{(\mathrm{LO})}
    &= i \boldsymbol{\Tc}^{\mathrm{(LO)}}_{1\rightarrow 3, \mathrm{PV}} \sum_{n=0}^{\infty}(\rho\,i\Kcs )^n
    \nn\\
    &= i \boldsymbol{\Tc}^{\mathrm{(LO)}}_{1\rightarrow 3, \mathrm{PV}}
    \left(1+
    \,\rho\,i\Mcs
    \right).
\label{eq:main_result}
\end{align}
Introducing $\boldsymbol{\mathcal{A}}_{1\rightarrow 3} \equiv \boldsymbol{\Tc}^{\mathrm{(LO)}}_{1\rightarrow 3, \mathrm{PV}}\,\Kcs^{-1}$, a short algebraic manipulation yields,
\begin{align}
      i\Tc_{1\to3}^{(\mathrm{LO})}
  &= \boldsymbol{\mathcal{A}}_{1\rightarrow 3}\,i\Mcs.
  \label{eq:main_result_v2}
\end{align}

This is Eq.~(\ref{eq:3B_main_v0}) for scenario 3. The full leading-order result for this scenario combines Eqs.~\eqref{eq:3B_main_v0} and \eqref{eq:case3}. To establish this second result we recall that $\boldsymbol{\Tc}^{\mathrm{(LO)}}_{1\rightarrow 3, \mathrm{PV}}$ includes 
 $\boldsymbol{\Tc}^{({\rm p})}_{1\to3}$, which is given in Eq.~\eqref{eq:T130_v1}.  This thus concludes the derivation of the leading-order expression for scenario 3.

 Because $\boldsymbol{\Tc}^{\mathrm{(LO)}}_{1\rightarrow 3, \mathrm{PV}}$ and $\Kcs$ are both real and non-singular, so is $\boldsymbol{\mathcal{A}}_{1\rightarrow 3}$. Consequently, the only phase of the amplitude comes from $\Mcs$, in accordance with the Watson-like theorem stated at the beginning of this section. The equivalent expression for $\Tc_{2\to 4}$, Eqs.~\eqref{eq:3B_main_v0} and \eqref{eq:case4v2} follows analogously by replacing the external probe with a hadron and replacing the off-shell form factor with an off-shell $\Mchoff$, which is then placed on shell by the nearby pole. We leave this as an exercise for the reader. 

We finish this section by remarking the key steps needed to derive $i\Tc_{2\to4}^{(\mathrm{LO})}$. This takes on the same form as the amplitude we have just considered, with the main difference being definition of the residue, which we can write in terms of $i\Ac^{h}_{2\rightarrow 4}$, defined in Eq.\eqref{eq:case4}. The steps needed to be taken towards this derivation are nearly identical to the ones taken to arrive at Eq.~\eqref{eq:main_result_v2}. There are two main conceptual differences. The first is the obvious dependence on $\Mch$, as opposed to the form factor of the heavy particle. This is made explicit in Eq.\eqref{eq:case4}. The second is a bit subtle. 

If we define the full $i\Tc_{2\to4}$ amplitude, this would be symmetric under the interchange of the three final $\varphi$ particles. That said, the leading-order contribution is due to a particular kinematic configuration, where the probe is carrying a much larger momentum than the other two $\varphi$ particles. One of the three $\varphi$ particles in the final state will then have approximately the same momentum as the probe. In this limit, and within the leading-order piece of the amplitude, we can identify this third particle as the $\varphi$ particle that scattered off the $\Phi$ in the target. There will be other pieces of the overall amplitude in which a fast $\varphi$ comes from another mechanism, but these are power-suppressed.
This implies that the leading-order contribution is an asymmetric component of the full amplitude. In other words, whether one chooses to symmetrize the full amplitude or not makes no difference in the conclusion drawn, as long as one picks out the piece of the amplitude that is leading order in quasi-free kinematics.

It is useful to summarize the steps taken to arrive at $\boldsymbol{\mathcal{A}}_{1\rightarrow 3}$ and its relation to the vertex function $\Gamma_3$ used in Eq.~\eqref{eq:case3}. First, $\boldsymbol{\Tc}^{\mathrm{(LO)}}_{1\rightarrow 3,\mathrm{PV}}$ absorbs, to all orders, all contributions in which the two-$\varphi$ loops are evaluated with the principal-value prescription. We then define $\boldsymbol{\mathcal{A}}_{1\rightarrow 3}\equiv\boldsymbol{\Tc}^{\mathrm{(LO)}}_{1\rightarrow 3,\mathrm{PV}}\Kcs^{-1}$, which allows the leading amplitude to be written as $i\Tc_{1\to3}^{(\mathrm{LO})}=\boldsymbol{\mathcal{A}}_{1\rightarrow 3}\,i\Mcs$. Finally, isolating the nearby $\Phi$ pole in $\boldsymbol{\mathcal{A}}_{1\rightarrow 3}$ gives the form displayed in Eq.~\eqref{eq:case3}, with $\Gamma_3$ defined by the residue of that pole. The non-relativistic matching in the following subsection identifies the product of $\Gamma_3\Kc_2$ and the nearly on-shell $\Phi$ propagator with the corresponding momentum-space bound-state wavefunction overlap.

\subsection{Non-relativistic limit - 3 body}
\label{sec:3body_LO_NR}

We now repeat the preceding construction for the breakup of a three-body bound state. Particles 1 and 2 are the two light particles, each of mass $m$, and particle 3 is the struck heavy particle, of mass $M$. For the final state we define,
\begin{equation}
   {\bf P}_{12}\equiv {\bf p}_1+ {\bf p}_2,
   \qquad
   {\bf p}_{\mathrm{rel}}\equiv \frac{{\bf p}_1- {\bf p}_2}{2},
   \qquad
   {\bf p}_{m}\equiv {\bf P}_{12}- \frac{2m}{2m+M}{\bf{P}}_i.
   \label{eq:momemtum:convention}
\end{equation}
where here, in contrast to Sec.~\ref{sec:2body_LO_NR}, ${\bf p}_{\rm rel}$ is the relative momentum between the two slow-moving $\varphi$ particles.
In the target rest frame, ${\bf p}_m={\bf p}_1+{\bf p}_2$ is the total momentum of the spectator pair. The struck heavy particle therefore carries momentum $-{\bf p}_m$ before the interaction with the probe and emerges with momentum ${\bf p}_3={\bf k}_{\Jc}-{\bf p}_m$.

In the quasi-free kinematics, the final state factorizes into an interacting two-body state of particles 1 and 2 and a separate one-body momentum state for particle 3. The non-relativistic limit of $\Tc_{1 \to 3}$ for the case of an external current probe then takes the form
\begin{align}
   \Tc_{1 \to 3}&=\sqrt{2M_b} \sqrt{2M} \sqrt{2m}^2 \langle \psi_{{\bf p}_1 {\bf p}_2 {\bf p}_3}^{(-)}|\Jc(0)|{\bf P}_i \rangle_{\rm NR}
   \label{eq:nrscenarioI3A}
   \\
   &=\sqrt{2M_b} \sqrt{2M} \sqrt{2m}^2 \langle \psi_{{\bf p}_{\rm rel}}^{(-)}|\langle -{\bf p}_m|\identity_1 \otimes \identity_2|\Psi_B^{\rm int} \rangle_{\rm NR} \frac{f(k_{\Jc}^2)}{2M} + \Tc_{1 \to 3}^{\rm ps}.
   \label{eq:nrscenarioI3B}
\end{align}
Here, $|\Psi_B^{\rm int} \rangle$ denotes a stationary state of the full three-body Hamiltonian in the three-body center-of-mass frame. To obtain this initial state we solve the three-body problem for a negative energy $-B$, $B>0$.
The state $|\psi_{{\bf p}_{\rm rel}}^{(+)} \rangle$ is the outgoing interacting two-body scattering state of the (12) subsystem, defined in its center-of-mass frame and labeled by the relative momentum ${\bf p}_{\rm rel}$.

It is therefore related to the plane wave $|{\bf p}_{\rm rel} \rangle$ by: 
\begin{equation}
    |\psi_{{\bf p}_{\rm rel}}^{(+)} \rangle=\bigg(1 + \hat{G}_0(E_{\rm rel}^{(+)})
    \hat{T} (E_{\rm rel}^{(+)}) \bigg)|{\bf p}_{\rm rel} \rangle,
    \label{eq:3Bscatteringwf}
\end{equation}
with $E_{\rm rel} \equiv {\bf p}_{\rm rel}^2/m$, since the two particles in the $(12)$ pair have equal masses, and $E^{(+)}_{\rm rel}=E_{\rm rel}+i \epsilon$.   

The first term on the right-hand side of Eq.~(\ref{eq:nrscenarioI3B}) is the leading-order contribution to the amplitude for breakup of the three-body bound state with a high-energy probe (for small ${\bf p}_m$). It equals the $\Tc_{1 \to 3}^{\rm (LO)}$ derived in the previous subsection within a relativistic formalism up to power-suppressed corrections. Note that the presence of the interacting $(12)$ final state $\langle \psi_{{\bf p}_{\rm rel}}^{(-)}|$ in this amplitude means it contains both the diagrams drawn explicitly in Fig.~\ref{fig:iT13_LO}: the plane-wave piece of the state gives the first diagram with no interaction between particles 1 and 2, and the $(12)$ FSI of the second diagram is computed using the $(12)$ T-matrix. When inserted in Eq.~(\ref{eq:nrscenarioI3B}) the $(12)$ FSI also contains the second piece of Eq.~(\ref{eq:main_result}) above, which is proportional to the on-shell $(12)$ scattering amplitude.

The factor of $\sqrt{2M_b} \sqrt{2M} \sqrt{2m}^2$ accounts for the use of non-relativistic, rather than relativistic, normalization for these three-body states, and allows this amplitude to be used in the standard relativistic relationship between the invariant amplitude squared and the differential cross section. The corresponding result for the non-relativistic scattering amplitude is:
\begin{equation}
\mathbf{T}_{1 \to 3}^{\rm NR}=(2 \pi)^4 \delta(E_f  - E_i - \omega_{\Jc}) \delta^{(3)}({\bf P}_f  - {\bf P}_i- {\bf k}_\Jc) \frac{f(k_\Jc^2)}{2M} \langle \psi_{{\bf p}_{\rm rel}}^{(-)}|\langle -{\bf p}_m|\identity_{\rm 1} \otimes \identity_{\rm 2}|\Psi_B^{\rm int} \rangle_{\rm NR} + \ldots,
\end{equation}
where $\ldots$ includes all power-suppressed contributions. Note that, as in the case of $\mathbf{T}_{1 \to 2}^{\rm NR}$ discussed in Sec.~\ref{sec:2body_LO_NR} above, the non-relativistic scattering amplitude includes overall energy- and momentum-conserving delta functions that are removed when computing $\Tc$.

\begin{figure}[t]
    \centering
    \includegraphics[width=.8\linewidth]{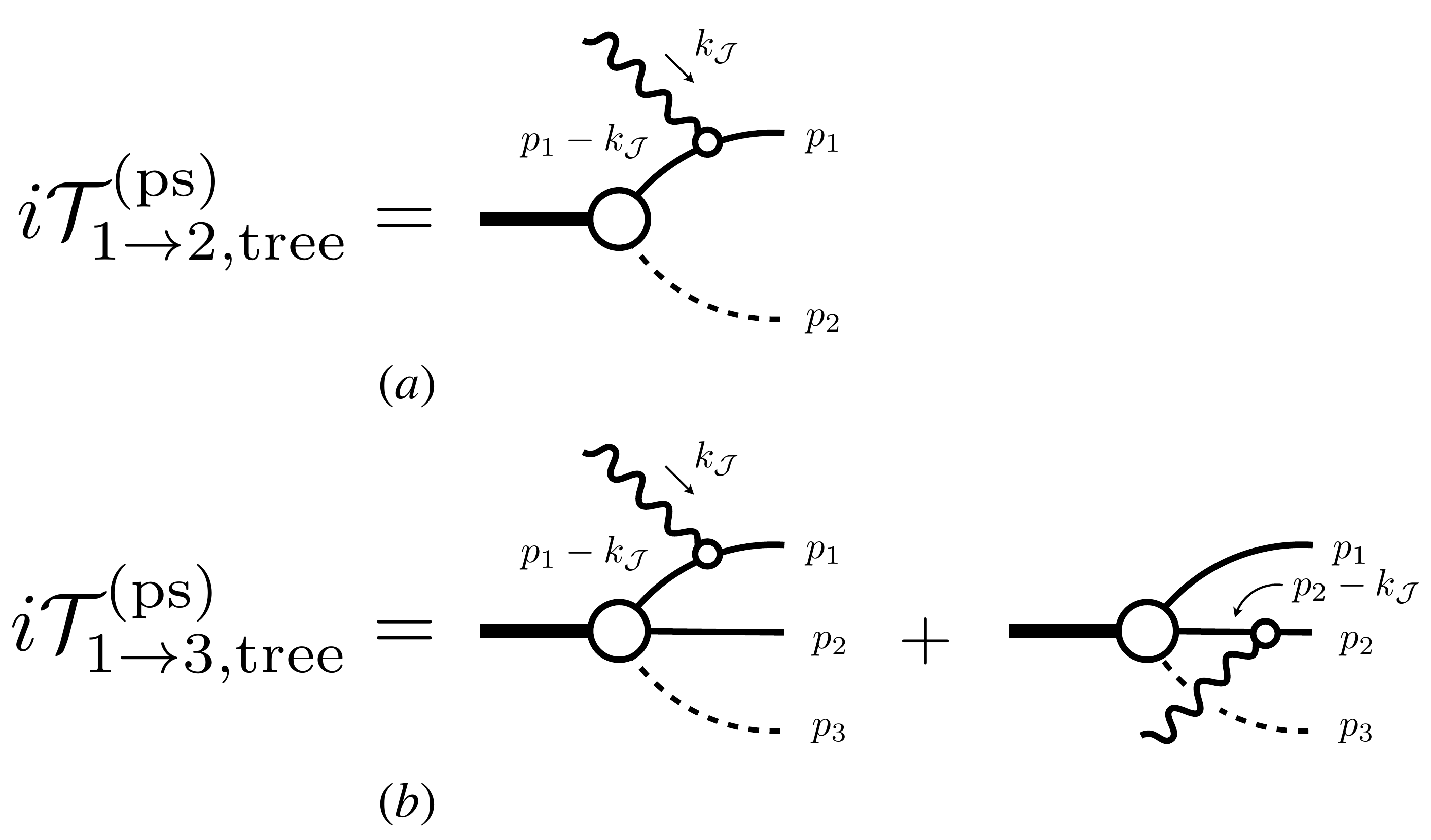}
    \caption{Diagrammatic representation of the tree-level power-suppressed contribution to ($a$) $i\Tc_{1\rightarrow 2}$  and ($b$) $i\Tc_{1\rightarrow 3}$.}
    \label{fig:tree_sub}
\end{figure}


\section{Subleading effects}
\label{sec:subleading}

We now turn to the remaining contributions appearing in the full amplitudes. The subleading contributions can be grouped into four distinct kinds:
\begin{enumerate}
    \item off-shell contributions to tree-level diagrams previously considered,
    \item subleading tree-level diagrams,
    \item loop corrections,
    \item poles associated with final-state rescattering.
\end{enumerate}
The goal is to show that each of these leads to kinematically suppressed contributions relative to the leading-order results already derived.

\textbf{1. Off-shell contributions to tree-level diagrams:}~The first class, defined as $\delta\Tc$ in Secs.~\ref{sec:LO_2Body} and \ref{sec:LO_3Body}, comprises the remainder of the tree-level contributions already considered. These do not have a nearby pole. By construction, these have the leading pole subtracted and are therefore suppressed in the kinematic region of interest. We focus our attention on the remaining three classes. Note that the arguments that follow are purely kinematic in nature and are therefore independent of the nature of the probe; they apply equally when the probe is a hadron. For simplicity, we phrase the arguments in terms of the current.

\textbf{2. Subleading tree-level diagrams:}~Next, we consider the other tree-level contributions, shown in Fig.~\ref{fig:tree_sub} for scenarios 1 and 3. These correspond to diagrams where the current couples to the slowly-moving particles in the final state.

To show that these are kinematically suppressed, it suffices to show that the propagator appearing in these diagrams has no nearby pole. We evaluate the denominator of the propagator carrying momentum $p_1-k_{\Jc}$,
\begin{align}
    (k_{\Jc}-p_1)^2-m^2
    &= p_1^2-m^2+k_{\Jc}^2 - 2k_{\Jc}\cdot p_1
\nn\\
 &= k_{\Jc}^2 - 2 k_{\Jc}\cdot p_1
\nn\\
 &\approx
 -{\bf k}_{\Jc}^2
-   {\bf k}_{\Jc}^2 \frac{\sqrt{m^2+{\bf p}_{1}^2}}{M}
+ 2 {\bf k}_{\Jc}\cdot {\bf p}_{1},
\end{align}
where in the last step we have dropped a contribution $\omega_{\Jc}^2\sim\mathcal{O}({\bf k}_{\Jc}^4/M^2)$. Using the momentum hierarchy of Eq.~\eqref{eq:Momenta_hiarchy}, which implies $|{\bf p}_{1}|\ll |{\bf k}_{\Jc}|$, and keeping only terms at zeroth order in ${\bf p}_{1}$, we find,
\begin{align}
    (k_{\Jc}-p_1)^2-m^2
 &\approx
 -{\bf k}_{\Jc}^2\left(1+ \frac{m}{M}\right).
\end{align}
This is of order $\mathcal{O}({\bf k}_{\Jc}^2)$, a large scale in the problem, and does not vanish. Consequently, these propagators have no nearby poles, and the resulting tree-level diagrams are kinematically suppressed compared to the leading-order contribution.

\begin{figure}[t]
    \centering
    \includegraphics[width=\linewidth]{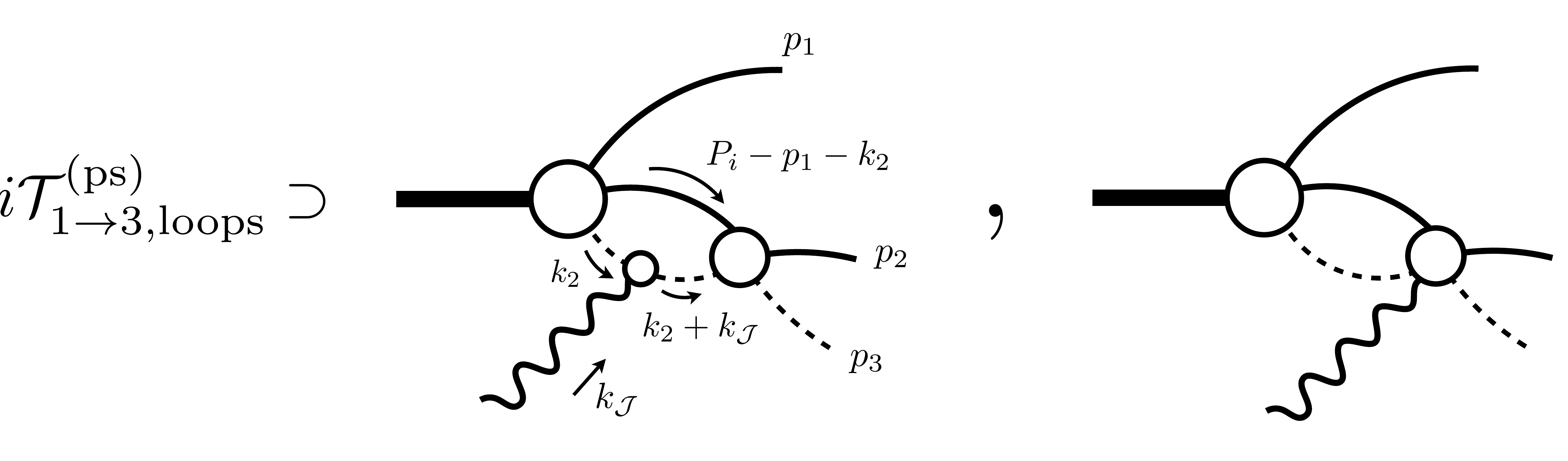}
    \caption{Examples of power-suppressed loop diagram contributions contained in $i\Tc_{1\rightarrow 3}^{(\mathrm{ps})}$.}
    \label{fig:loops_sub}
\end{figure}

\textbf{3. Loop contributions \`a la Landau:}~Loop diagrams generically soften singularities~\cite{Landau:1959fi}, so we expect their contributions to be kinematically suppressed. Here we give a simple explanation of why this is the case, without deriving a closed form for the full amplitude. Since we are only interested in showing that these singularities are softer than a pole, this level of argument is sufficient. Related analyses of the interplay between Landau singularities and final-state rescattering in multi-hadron amplitudes can be found in Ref.~\cite{Sakthivasan:2024uwd}, where it was shown that rescattering corrections generally preserve the location of the underlying triangle singularity while softening its contribution to the full amplitude.  

For simple loops, the suppression is immediately evident. In Fig.~\ref{fig:loops_sub}, we show two examples of loop diagrams contributing to $\Tc_{1\to3}$. The first is a triangle diagram of the type studied in Ref.~\cite{Briceno:2020vgp}, which has at most a logarithmic singularity associated with all particles simultaneously going on-shell.

Since the initial hadron is a bound state, its mass satisfies $P_i^2 < (2m+M)^2$. This means that the three internal lines carrying momenta $p_1$, $k_2$, and $P_i-p_1-k_2$ cannot all simultaneously go on shell. Since the line with momentum $p_1$ is by definition on shell, the remaining two cannot go on shell simultaneously. Hence at most two of the three propagators inside the loop can go on shell, implying at most a square-root singularity of the kind discussed in Sec.~\ref{sec:LO_3Body}. For loosely bound initial states, the particles in the loop are nearly on shell, and these diagrams have nearby logarithmic singularities of the form $\log\left((s_{23}-s_{\rm trig.})/m^2\right)$, where $s_{23}=(p_2+p_3)^2$ and $s_{\rm trig.}$ is a kinematic variable depending on, among other things, $k_\Jc$. Since not all particles inside the loop can go on shell, $s_{23}-s_{\rm trig.}$ never vanishes in the physical kinematic region.

The second diagram in Fig.~\ref{fig:loops_sub} depicts a short-distance contribution of the current, typically described as a two-body current contribution. This diagram has no singularities, since the particles in the loop cannot go on shell. If the loop appeared after the two-body current insertion, it would generate a phase-space singularity of the kind discussed in Sec.~\ref{sec:LO_3Body}. This shows that higher-body contributions to the current matrix element are suppressed relative to the one-body contribution. 

\begin{figure}[t]
    \centering
    \includegraphics[width=1\linewidth]{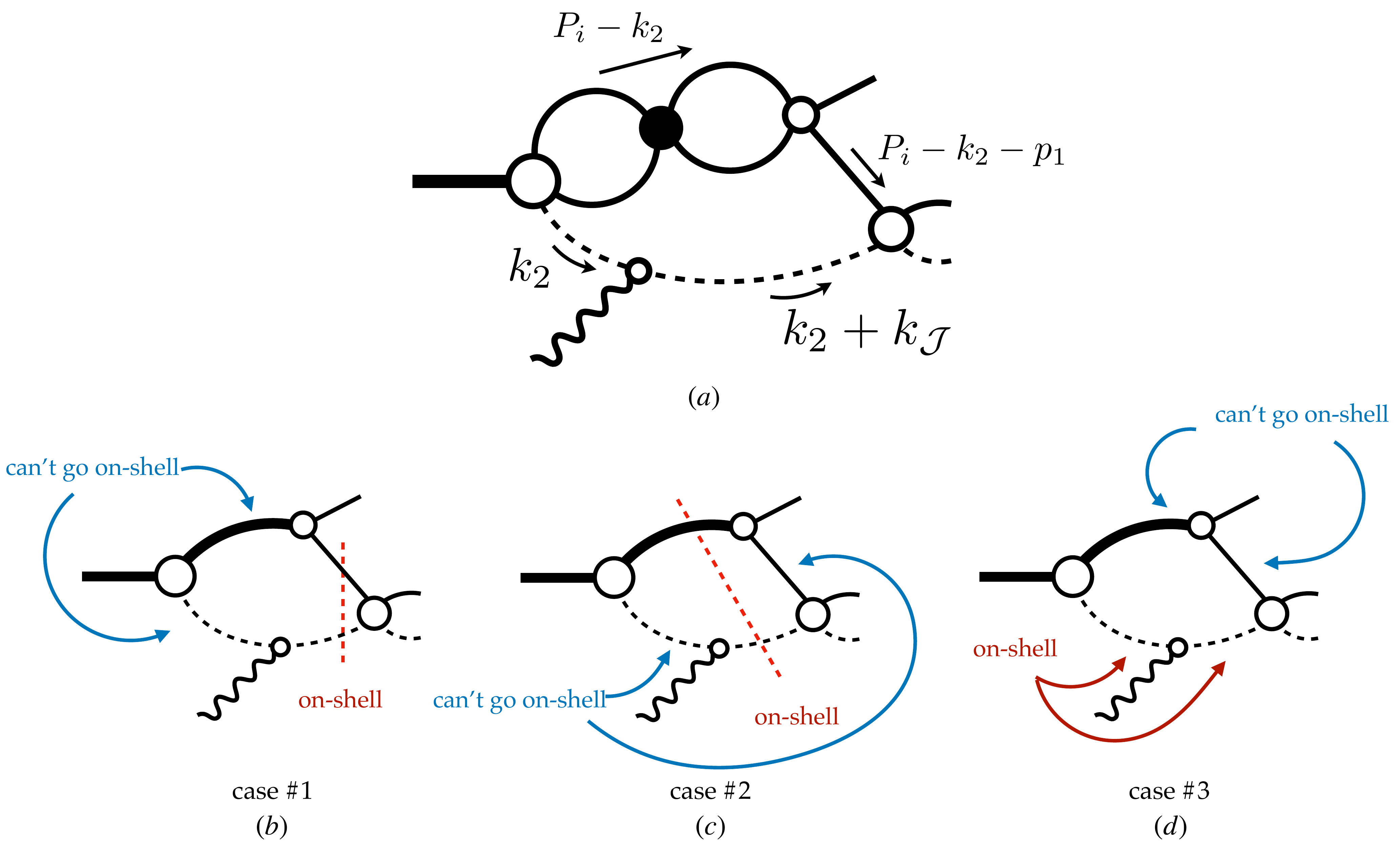}
    \caption{($a$) A loop diagram that, for a strongly interacting $2\varphi$ subsystem, could naively be expected to have box singularities. ($b$)--($d$) make the box-like nature explicit by replacing the off-shell $\Mcs$ with a simple propagator. In each case, we explain why at most two internal propagators can go on shell, which is sufficient to show that no box singularity is present. }
    \label{fig:box}
\end{figure}

Less trivial cases arise when the system is strongly interacting and supports nearby resonances or bound states, as is relevant for nuclear systems. In such cases, a subsystem of the final-state particles can interact strongly to form a bound state or resonance before rescattering with another particle. For scenario 3 in particular, this is illustrated in Fig.~\ref{fig:box}($a$). If the two-$\varphi$ scattering amplitude $\Mcs$ has a nearby pole due to a bound state or resonance, the loop in the figure more closely resembles a box diagram than a triangle. Near the pole, $\Mcs$ behaves as a single-particle propagator, $\Mcs((P_i-k_2)^2)\sim-g^2_0/[(P_i-k_2)^2-s_0]$, where $s_0$ is the location of the pole. The lower row of Fig.~\ref{fig:box} makes this box-like structure explicit by replacing the off-shell $\Mcs$ with this simple propagator.

Box diagrams have singularities scaling as $1/\sqrt{s-s_{\rm box}}$~\cite{Landau:1959fi}, which more closely resemble a pole. These singularities occur when all four intermediate particles go on-shell. Since resonances correspond to poles in the complex plane, they cannot go on-shell for real energies. We therefore focus on the most potentially problematic case: a shallow real bound state of mass $m_0=\sqrt{s_0}<2m$, which we refer to as a dimer and denote $d$.

Since the initial hadron is stable, its mass satisfies $\sqrt{P_i^2}<m_0+M$. We now argue that this kinematic constraint ensures that at most two propagators in the box diagrams can simultaneously go on-shell. As a result, the most severe box singularities are absent, and the nearest singularities are logarithmic, just as in the triangle case.

Consider the three box diagrams in Figs.~\ref{fig:box}($b$)--($d$). The leftmost vertex in each is the $H_3\to d+\Phi$ vertex. Since this decay is kinematically forbidden, at most one of the $d$ or $\Phi$ propagators can go on shell inside the loop; if both could, the kinematics of $P_i$ would permit the physical decay $H_3\to d+\Phi$.

In Fig.~\ref{fig:box}($b$), neither the $d$ nor the $\Phi$ propagators go on-shell. This immediately excludes the most severe box singularity. The actual singularity arises from the final $\Phi$ and the exchanged $\varphi$ going on shell, as indicated by the dashed cut lines.

In Fig.~\ref{fig:box}($c$), we allow the $d$ to go on-shell. This forbids the $\Phi$ propagator with momentum $k_2$ from going on-shell. Furthermore, since $d$ is a bound state of two $\varphi$ particles and the external legs are on shell, the exchanged $\varphi$ cannot go on shell either. Hence, again, at most two propagators go on-shell.

In Fig.~\ref{fig:box}($d$), we allow the $\Phi$ with momentum $k_2$ to go on-shell. For this to occur, the momentum flowing through the $d$ propagator must satisfy $(P_i-k_2)^2<m_0^2<(2m)^2$, which prevents the $d$ from going on-shell. The external on-shell condition also prevents the exchanged $\varphi$ from going on-shell. The $\Phi$ propagator with momentum $k_2+k_\Jc$ is not forbidden from going on shell, but this does not produce a square-root singularity, since no $s$-channel cut is present.

In general, the nearest singularity is logarithmic rather than box-like. This follows from the fact that the dimer is bound, so there is an energy gap between the $d$ and the two-$\varphi$ threshold. Consequently, the $d$ and the exchanged $\varphi$ can never simultaneously go on shell, regardless of the kinematics.

\begin{figure}[t]
    \centering
    \includegraphics[width=.9\linewidth]{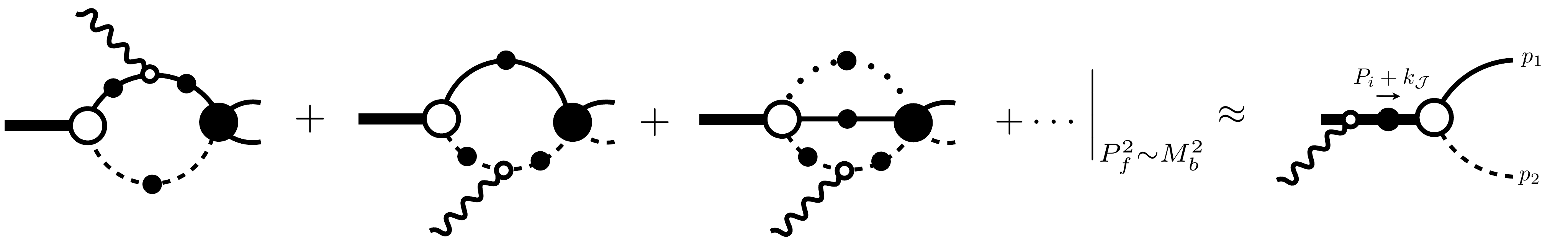}
    \caption{Shown is the bound-state pole contribution in scenario 1.}
    \label{fig:bs_pole}
\end{figure}

\textbf{4. Bound-state pole contributions:}~There is one further class of pole contributions to consider, arising from final-state interactions in which the system recombines into a bound state. This is depicted in Fig.~\ref{fig:bs_pole} for scenario 1. Although this requires loops, it is sufficiently distinct from the loop contributions discussed above that we treat it separately.

The final state couples to an off-shell pole of the form $\sim[(k_{\Jc}+P_i)^2-M_b^2]$, where $M_b$ is the bound-state mass. 
Applying the same approximations as for the tree-level diagrams above, the denominator of this pole behaves as,
\begin{align}
    (k_{\Jc}+P_i)^2-M_b^2
    &= P_i^2-M_b^2+k_{\Jc}^2 + 2k_{\Jc}\cdot P_i
\nn\\
 &= k_{\Jc}^2 + 2 k_{\Jc}\cdot P_i
\nn\\
 &\approx
 -{\bf k}_{\Jc}^2
+   {\bf k}_{\Jc}^2 \frac{M_b}{M}
\nn\\
 &= {\bf k}_{\Jc}^2 \left( \frac{M_b}{M} - 1\right).
\end{align}
This is again of order $\mathcal{O}({\bf k}_{\Jc}^2)$ and does not vanish, confirming that these poles are far from the kinematic region of interest.

Having presented a general argument as to why all other contributions are kinematically suppressed, we proceed to explicitly calculate some of the power-suppressed contributions for $\Tc_{1 \to 2}$ and $\Tc_{1 \to 3}$ using non-relativistic quantum mechanics. This exercise serves two purposes. On one hand, we will gain confidence in the arguments above. On the other hand, we will gain further insight into the kinematic behavior at low energies of the power-suppressed corrections. We begin by considering the corrections to $\Tc_{1 \to 2}$. In this case, the corrections can be understood as arising from final state interactions after the heavy particle is struck by the probe. 

\subsection{Non-relativistic analysis of suppressed diagrams: two-body breakup }
\label{sec:NR_loops_2B}

\begin{figure}[t]
    \centering
    \includegraphics[width=.7\linewidth]{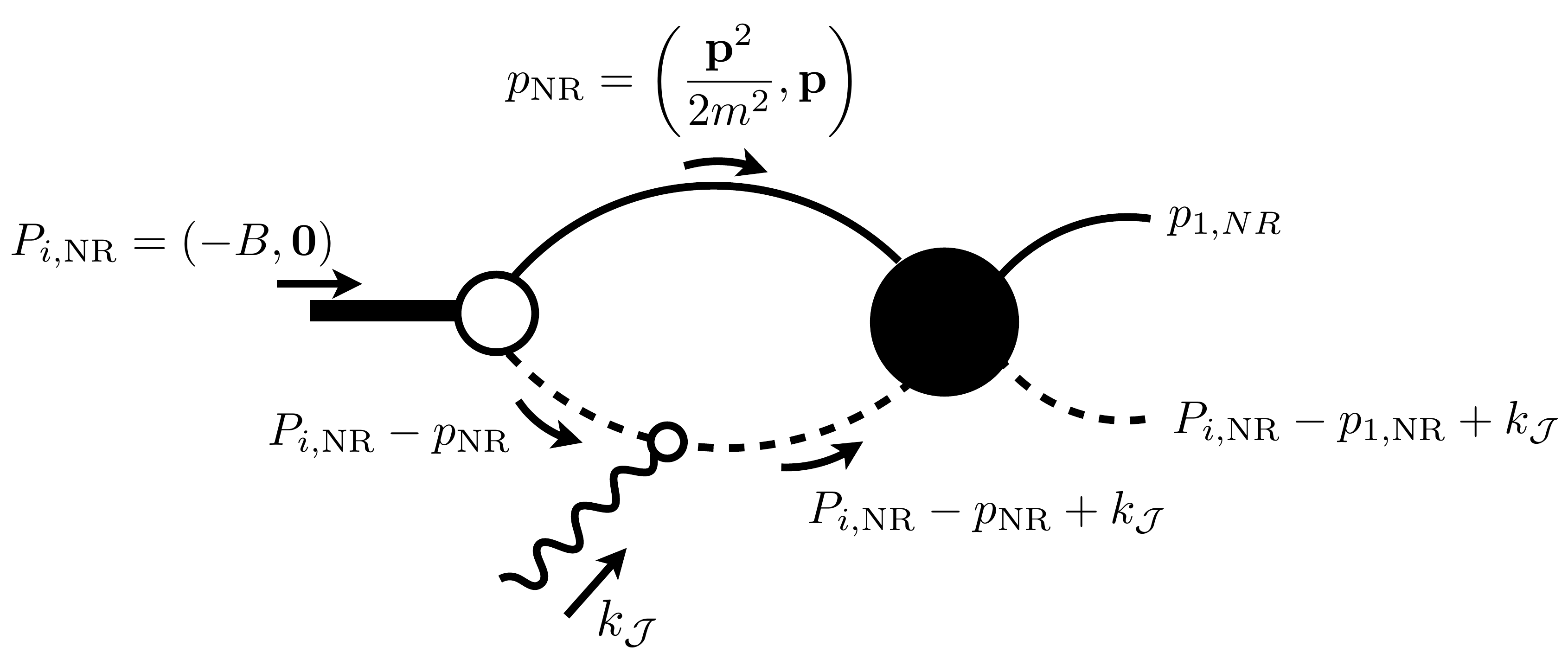}
    \caption{ Shown is a class of subleading diagrams where the current couples to an internal particle of type 2.  }
    \label{fig:NR_1body_loop}
\end{figure}

Corrections involving final-state interactions appear in $\Tc_{1 \to 2}^{\rm (ps)}$. There are three main contributions that constitute final-state interactions. The first are due to particle 2 being struck, followed by the two particles interacting, see Fig.~\ref{fig:NR_1body_loop}. The second is due to particle 1 being struck, followed by the two particles interacting. 
Meanwhile, there are also two-body-current contributions, which are always power suppressed and which we do not discuss explicitly here. Because we are considering kinematics where particle $2$ is the fast-moving particle, one can argue that the dominant of these three contributions is the first.

We will denote the FSI contribution $\Tc_{1 \to 2}^{\rm ps,FSI}$. Because unitary transformations can reshuffle non-singular contributions between this term and those involving two-body currents, defining a specific final-state interaction contribution only makes sense within a particular field representation. However, once a set of fields, and hence a set of two-body interactions, has been defined, the FSI contribution is given by:
\begin{eqnarray}
    \Tc_{1 \to 2}^{\rm ps,FSI}=\sqrt{2M_b}\sqrt{ 2m}\sqrt{2M} \int \frac{d^3 p}{(2 \pi)^3} \, \langle {\bf p}_{\rm rel}|\hat{T}\left(E_{\mathrm{rel}}^{(+)}\right)|{\bf p} - \frac{m_R
    }{M} {\bf k}_\Jc\rangle_{\rm NR} \nonumber\\
   \qquad \qquad  \frac{1}{\omega^{(+)}_\Jc-B-\frac{(-{\bf p}+{\bf k}_\Jc)^2}{2M} -\frac{{\bf p}^2}{2m}}\frac{f({\bf k}_\Jc^2)}{2M}\langle {\bf p}|\Psi_B^{\mathrm{int}} \rangle_{\rm NR},
\end{eqnarray}
where $E_{\mathrm{rel}}=\omega_\Jc-B-\frac{{\bf k}_\Jc^2}{2(M+m)}$ and
$\hat{T}$ is the two-body $\varphi \Phi$ T-matrix. The outgoing scattering state is related to the plane-wave state by the Lippmann–Schwinger equation, $\langle\psi_{{\bf p}_{\rm rel}}^{(-)}| = \langle {\bf p}_{\rm rel}|+\langle {\bf p}_{\rm rel}|\hat{T}(E_{\mathrm{rel}}^{(+)})\hat{G}_0(E_{\mathrm{rel}}^{(+)}) $, so that the second term encodes final-state interactions, while the first one leads to the impulse approximation of Sec.~\ref{sec:2body_LO_NR}.
As in Sec.~\ref{sec:2body_LO_NR}, the states are eigenstates of the relative-momentum operator, normalized according to Eq.~(\ref{eq:NRnormn}).
Note that the on-shell singularity then occurs for $|{\bf p}_{\rm rel}|=|-\frac{m_R}{M}{\bf k}_\Jc+{\bf p}|$. Hence we parameterize ${\bf p}_{\rm rel}=-\frac{m_R}{M}  {\bf k}_\Jc+{\bf p}_m \equiv {\bf q}_{\mathrm{on}}$, where ${\bf p}_m$ is the missing momentum. 
If we then substitute for $\omega_\Jc - B$ from Eq.~(\ref{eq:omega}) we can rewrite the intermediate-state propagator as:
\begin{equation}
    \frac{1}{\frac{{\bf k}_\Jc^2}{2(M+m)} + \frac{(\frac{m_R}{M}{\bf k}_\Jc-{\bf p}_m)^2}{2 m_R}-\frac{(-{\bf p}+{\bf k}_\Jc)^2}{2M} -\frac{{\bf p}^2}{2m} + i \epsilon} \approx  \frac{M}{{\bf k}_\Jc \cdot ({\bf p}-{\bf p}_m) + i \epsilon} .
    \label{eq:eikonal}
\end{equation}
Note that, prior to any approximations, the terms proportional to ${\bf k}_{\Jc}^2$ cancel exactly. The denominator then reduces to $\frac{{\bf k}_{\Jc}\cdot({\bf p}-{\bf p}_m)}{M}+\frac{{\bf p}_m^2-{\bf p}^2}{2m_R}+i\epsilon$. In the region of integration where ${\bf p}$ and ${\bf p}_m$ are much smaller than ${\bf k}_\Jc$, the quadratic terms in the momenta are suppressed relative to the linear term and can be neglected. The propagator therefore reduces to the eikonal form given in Eq.~(\ref{eq:eikonal}).

We are now in a position to estimate the size of the power-suppressed corrections from this region of the integration. From the eikonal denominator in Eq.~(\ref{eq:eikonal}), the pole condition fixes the component of ${\bf p}$ parallel to ${\bf k}_\Jc$ to be close to the corresponding component of ${\bf p}_m$. The remaining transverse support is set by the bound-state wavefunction.
The soft contribution to the integral is therefore of order
\begin{equation}
    [\Tc_{1 \to 2}^{\rm ps, FSI}]_{p \sim p_m} \sim \frac{p_m}{k_\Jc} p_m \langle {\bf q}_{\mathrm{on}}|\hat{T}\left(E_{\mathrm{rel}}^{(+)}\right)|{\bf q}_{\mathrm{on}}\rangle_{\rm NR} \Tc_{1 \to 2}^{\rm LO }.
\end{equation}
The first factor indicates the eikonal suppression from the longitudinal integration, and the second the transverse phase-space scale.
If $s$-waves dominate the final-state interaction, the on-shell amplitude takes the form:
\begin{equation}
    \langle {\bf q}_{\mathrm{on}}|\hat{T}\left(E_{\mathrm{rel}}\right)|{\bf q}_{\mathrm{on}}\rangle_{\rm NR} = \frac{2 \pi}{m_R |{\bf q}_{\mathrm{on}}|} e^{i \delta(E_{\mathrm{rel}})} \sin\delta(E_{\mathrm{rel}}).
\end{equation}

For $|{\bf k}_\Jc|\gg |{\bf p}_m|$, one has ${\bf q}_{\mathrm{on}}\sim \frac{m_R}{M}k_{\Jc}$, so the final state interactions are suppressed by an additional power of order $1/k_{\Jc}$. Combining this with the eikonal suppression from the longitudinal integration and the soft transverse phase-space scale, this gives

\begin{equation}
  [\Tc_{1 \to 2}^{\rm ps, FSI}]_{p \sim p_m} \sim 
  \frac{p_m^2}{k_\Jc^2} \sin\delta(E_{\mathrm{rel}}),
\end{equation}
and is therefore further suppressed when the phase-shift is small.

We must now consider the contribution of other regions in the ${\bf p}$ integral to $\Tc_{1 \to 2}^{\rm ps, FSI}$. For $|{\bf p}| \sim |{\bf k}_\Jc|$ the amplitude is to be evaluated in off-shell kinematics. In this region, the propagator is of order $m/k_{\Jc}^2$ and the wave function is evaluated for momenta of order ${\bf k}_{\Jc}$. The extent to which this contribution is suppressed therefore, depends on the behavior of wave functions (and amplitudes) that are off-shell by an amount of order $k_{\Jc}$. The slowest possible fall-off of wave functions is found in a theory with only short-range interactions, in which case 
\begin{equation}
\left |\frac{\psi({\bf k}_\Jc)}{\psi({\bf p}_m)}\right| \sim \frac{p_m^2}{k_\Jc^2}.
\end{equation}
If we also make the minimal assumption that the size of the amplitude in this off-shell region is $\sim \frac{1}{m_R k_\Jc}$ (i.e. it is order 1 times its normalization), then we have a size for the contribution of this region to the integral of 
\begin{equation}
   [\Tc_{1 \to 2}^{\rm ps, FSI}]_{p \sim k_\Jc} \sim k_\Jc^3 \frac{1}{m_R k_\Jc} \frac{m_R}{k_\Jc^2} \frac{p_m^2}{k_\Jc^2}\Tc_{1 \to 2}^{\rm LO}. 
\end{equation}

This is parametrically of the same order as the contribution from modes with ${\bf p} \sim {\bf p}_m$. The relative importance of the two regions depends on the behavior of the phase shifts at high energy and the falloff of the two-body wavefunction at momenta of order $k_\Jc$. Nevertheless, independently of these details, we conclude that $\Tc_{1 \to 2}^{\rm ps, FSI}$ is suppressed by at least a factor of $p_m^2/k_\Jc^2$ compared to the leading contribution of the amplitude.

\subsection{Non-relativistic analysis of suppressed diagrams: three-body breakup }
\label{sec:NR_loops_3B}

We can pursue a similar analysis of suppressed final-state interactions in the breakup of the three-body bound-state. Here, though, we need to use the multiple-scattering series to express the three-body $T$-matrix $\hat{T}_{123}$, which appears in the three-body version of Eq.~(\ref{eq:3Bscatteringwf}), in terms of two-body amplitudes. 

First, we note that the (12) FSI appearing in the second diagram of Fig.~\ref{fig:iT13_LO} is already included as part of the leading amplitude (\ref{eq:nrscenarioI3B}). As in the two-body case, among the power-suppressed contributions, the dominant term arises from final-state interactions between the fast (struck) particle 3 and one of the soft (unstruck) particles 1 and 2. These are depicted in Fig.~\ref{fig:loops_sub}. 
Once a set of fields, and hence a set of two-body interactions, has been specified, these contributions can be computed systematically. 

Consider, for example, the correction in which particle 2 rescatters from particle 3 after the current insertion. Following the momentum routing of Fig.~\ref{fig:loops_sub}, let ${\bf k}_2$ denote the momentum of particle 3 prior to the current insertion. In the target rest frame, momentum conservation implies that the total momentum of the two slow particles is $-{\bf k}_2$, and we therefore define ${\bf q} \equiv -{\bf k}_2$. With this choice, ${\bf q}$ represents the total momentum of the two slow particles before the (23) hard scattering and we can write this FSI correction as:

\begin{eqnarray}
\Tc_{1 \to 3}^{\rm ps, (1)}
   &=&\sqrt{2M_b} \sqrt{2M} \sqrt{2m}^2 \int \frac{d^3q}{(2 \pi)^3} \langle {\bf a}_f|\hat{T}_{23}(E_{23}^{(+)})|{\bf a}_i({\bf q}) \rangle_{\rm NR} \nonumber\\
      &&  \qquad \qquad \qquad\times\frac{1}{ \omega_\Jc^{(+)} -B- \frac{{{\bf p}_1}^2}{2m} - \frac{({\bf q} - {{\bf p}_1})^2}{2m} - \frac{(  {\bf k}_{\Jc}-{\bf q} )^2}{2M}}
 \times \Psi_B^{\rm int}({\bf p}_1-\frac{{\bf q}}{2} ,{\bf q})  \frac{f(k_{\Jc}^2)}{2M},
   \label{eq:3BhighenergyFSI}
\end{eqnarray}
where $\hat{T}_{23}$ is the two-body $\varphi\Phi$ scattering matrix and
\begin{equation}
    {\bf a}_f={\bf p}_m-\frac{m_{R}}{m}{\bf p}_1-\frac{m_{R}}{M}{\bf k}_{\Jc},
    \qquad
    {\bf a}_i({\bf q})={\bf q}-\frac{m_{R}}{m}{\bf p}_1-\frac{m_{R}}{M}{\bf k}_{\Jc}.
\end{equation}
In Eq.~(\ref{eq:3BhighenergyFSI}), $E_{23}$ denotes the relative energy of the (23) subsystem, which in terms of the on-shell relative momentum ${\bf a}_f $ of particles 2 and 3 is given by $E_{23} = {\bf a}_f^2/(2m_R)$, with $m_R$ the reduced mass of the (23) subsystem. The wavefunction $\Psi_B^{\rm int}({\bf p},{\bf q})$ is the three-body wave function expressed in terms of the Jacobi variables:  ${\bf p}$ is the relative momentum of the slow particles and ${\bf q}$ is their total momentum in the rest frame of the initial bound-state.

Energy conservation for the physical final state together with the missing momentum convention given in Eq.~(\ref{eq:momemtum:convention}) gives
\begin{equation}
    \omega_\Jc-B = \frac{({\bf p}_1)^2}{2m}
    +\frac{({\bf p}_m-{\bf p}_1)^2}{2m}
    +\frac{({\bf k}_{\Jc}-{\bf p}_m)^2}{2M}.
\end{equation}
Substituting this back into the propagator appearing in Eq.~(\ref{eq:3BhighenergyFSI}) and assuming quasi-free kinematics, in which the momenta of the spectator particles satisfy $|{\bf q}|, |{\bf p}_m|, |{\bf p}_1| \ll |k_\Jc|$, we obtain the eikonal form of the propagator:
\begin{equation}
    \frac{1}{ \omega_\Jc^{(+)} -B- \frac{{{\bf p}_1}^2}{2m} - \frac{({\bf q} - {{\bf p}_1})^2}{2m} - \frac{(  {\bf k}_{\Jc}-{\bf q})^2}{2M}} \approx \frac{M}{{\bf k}_{\Jc}\cdot({\bf q}-{\bf p}_m)+i\epsilon}.
\end{equation}
Thus, like in the two-body case, the loop is dominated by ${\bf q}\sim{\bf p}_m$. In this region, the relative momentum of the (23) subsystem is ${\bf a}_f \sim {\bf a}_i({\bf q})\approx -\frac{m_R}{M}{\bf k}_\Jc$, and thus, the two-body $T-$matrix is being evaluated at a momentum of this order and scales as $ 1/(m_R {\bf k}_\Jc)$. So, as in the two-body case, the FSI integral is suppressed by two powers of ${\bf p}_m/{\bf k}_\Jc$ compared to the leading amplitude:
\begin{equation}
    \Tc_{1 \to 3}^{\rm ps, (1)} \sim p_m^3 \frac{1}{k_{\Jc}} \frac{1}{p_m k_{\Jc}} \Tc_{1 \to 3}^{\rm LO} \sim \frac{p_m^2}{k_{\Jc}^2} \Tc_{1 \to 3}^{\rm LO},
\end{equation}
with additional suppression occurring if the interaction between particles 2 and 3 is such that the corresponding phase shifts are small at energies of order $\omega_\Jc$. Two-body currents, as depicted in the second diagram of Fig.~\ref{fig:loops_sub}, can produce an effect that is of the same size, but not larger.

Additional soft rescatterings within the (12) subsystem after the hard 23 rescattering do not remove the suppression found above. These interactions are governed by the low-energy (12) amplitude and can modify the result by a factor of order $\frac{p}{ip+1/a}$, where $p$ is the relative loop momentum in the (12) subsystem. This factor can be important when the (12) interaction is strong, but it does not introduce powers of the hard scale ${\bf k}_{\Jc}$. Therefore the contribution remains suppressed by
\begin{equation}
   \Tc_{1 \to 3}^{\rm ps, (FSI)} \sim \sim \frac{p_m^2}{k_{\Jc}^2} \frac{p}{1/a + ip} \Tc_{1 \to 3}^{\rm LO}.
\end{equation}
%

\section{Summary and outlook}
\label{sec:summary}

In this work, we have derived the leading-order contributions to the amplitude for the fast breakup of a weakly bound two- or three-body bound state by a hard external probe. We have shown that in kinematics in which the probe is collinear with the struck particle, and so the struck particle is approximately on-shell before the collision, all other contributions are suppressed by powers of $p_m/k_\Jc$, where $p_m$ is the missing momentum and $k_\Jc$ is the momentum transfer. For the two-body case, the leading-order amplitude reduces to the impulse approximation: the probe couples to a single constituent and the spectator is left behind with small momentum. For the three-body case, we have proven a Watson-like theorem: the phase of the leading-order amplitude is entirely determined by the on-shell scattering amplitude of the low-energy two-body subsystem not struck by the probe. These arguments are purely kinematic in nature and hold regardless of whether the probe is a current or a hadron. Subleading effects include off-shell corrections to tree-level diagrams, tree-level diagrams involving two-body currents, loop corrections in which the slow-moving particles are not on shell, loop corrections in which one or more slow-moving particles scatters off the struck, fast-moving, particle, and bound-state poles from final-state rescattering. 

The main results have been derived in two complementary frameworks: a fully relativistic quantum field theory and a non-relativistic effective theory. The relativistic derivation establishes the results without dynamical assumptions, while the non-relativistic treatment makes the connection to the existing nuclear literature more transparent and invokes naive dimensional analysis for loop corrections to estimate the size of the subleading corrections. The agreement between the two frameworks confirms that the leading-order results and their kinematic suppression are independent of whether one works with relativistic or non-relativistic kinematics. 

To further illustrate the general statements of the main text in a concrete setting, we have included appendix \ref{app:EFT_example} in which the framework is applied to halo EFT~\cite{Hammer:2017tjm}, the effective theory describing one-neutron halo nuclei. There, we explicitly compute both the leading impulse approximation contribution and the power-suppressed final-state interaction and two-body current corrections, confirming the general suppression pattern derived in the main text.

Several generalizations of this work are worth pursuing. First, our proof of the factorization of $\Tc_{1 \to 3}$, and the appearance of the two-body phase in the soft part of the result, assumed that only short-range forces were operative. The proof can be generalized to the situation where QED interactions are also present. If two charged particles are left behind, then the low-energy scattering amplitude that appears in the factorized result will include the Rutherford amplitude. Coulomb phases will also then appear in the short-range piece of the amplitude we focused on here. Factorization still holds, even if a charged core is removed and some of the left-behind constituents are charged, as long as the Sommerfeld parameter associated with the fast-moving, $\eta_{\rm fast}=Z_{\rm soft} Z_{H} \alpha_{\rm QED}/v_{\rm rel}$,  core is small. Non-Coulombic QED corrections are suppressed by powers of $v/c$ or powers of $\alpha_{\rm QED}$.

Second, while we have focused on scenarios where at most two particles are left behind after the probe interaction, the framework can be extended to the case where $N$ particles are left behind. In this case, the final-state interactions among the remaining particles are described by the $N$-body scattering amplitude. This is also of relevance for placing constraints on unparticle systems~\cite{Hammer:2021zxb}, where the number of constituents is not well-defined, and for studying the internal structure of exotic hadrons. Such a proof can be carried out straightforwardly to the case where \emph{three bodies are left behind} by making use of ongoing efforts aimed at deriving the constraints from unitarity for three-hadron systems~\cite{Hansen:2015zga,Briceno:2019muc,Jackura:2019bmu,Jackura:2022gib,Draper:2023xvu, Briceno:2024ehy, Feng:2024wyg,Sadasivan:2021emk,Mai:2017wdv,Mikhasenko:2019vhk}.
The energy spectrum of the four neutrons left behind in the  hard knock-out reaction $^8$He$(p,p\alpha)4n$ was calculated 
in Ref.~\cite{Lazauskas:2022mvq} in a model calculation based on the sudden approximation.
An extension of our proof to four bodies left behind would rigorously justify this approximation. 

Third, throughout this work, we have restricted attention to scalar particles, thereby avoiding complications associated with spin. Extending the framework to include particles with spin is conceptually straightforward: the main effect is to introduce additional angular structure in the amplitudes, governed by the spin of the constituents and the probe. The leading-order results are expected to retain the same kinematic suppression structure, with the spin degrees of freedom entering through additional form factors and coupling constants that can be determined from independent measurements or EFT power counting.

The advent of radioactive-beam facilities that can produce fast beams of neutron-rich nuclei provides new opportunities to examine multi-neutron systems that are left behind upon fast-particle removal in quasi-free kinematics. The results proved here facilitate the controlled extraction of information on multi-neutron interactions from data on such reactions, since they delineate the size and kinematic dependence of corrections to the leading, quasi-free amplitude. This approach is currently being used to extract the case of the neutron-neutron scattering length from data on $\alpha$ particle knockout from ${}^6$He~\cite{RIKENnnProp}.

\section*{Acknowledgments}
We thank the Institute for Nuclear Theory at the University of Washington and the organizers of the program Quantum Few- and Many-Body Systems in Universal Regimes (24-3), where this work was initiated, for their kind hospitality and stimulating research environment. We thank Matthias G\"obel for feedback on an earlier version of the manuscript. This research was supported in part by the INT's U.S. Department of Energy grant No. DE-FG02- 00ER41132. RAB and DRP were partly supported by the U.S. Department of Energy, Office of Science, Office of Nuclear Physics under Award Nos. DE-AC02-05CH11231 (RAB) and DE-FG02-93ER40756 (DRP).
H.-W.H. was supported in part by the Deutsche Forschungsgemeinschaft (DFG, German Research Foundation), Project ID No. 279384907—SFB 1245, and by the German Federal Ministry of Research, Technology and Space (BMFTR), Contract No. 05P24RDB.

\appendix

\bibliography{bibi.bib}

\section{EFT Example: Fast breakup of a one-neutron halo nucleus}
\label{app:EFT_example}

To illustrate the general statements in the main text, we explicitly consider the fast breakup of a one-neutron $S$-wave halo nucleus with a spin-0 core by an external current $\Jc$. The system is described by the following non-relativistic effective Lagrangian~\cite{Hammer:2017tjm},
\begin{eqnarray}
\mathcal{L} &=&
n^\dagger \left( i\partial_0 + \frac{\nabla^2}{2m} \right) n
+ c^\dagger \left( i\partial_0 + \frac{\nabla^2}{2M} \right) c+
 \Delta h^{\dagger} h
- g
\left( h^{\dagger}nc  + {\rm h.c.} \right)+\ldots\,,
\label{eq:ncL}
\end{eqnarray}
where $c$ is the spinless core field with mass $M$, $n$ is the neutron field with spin $1/2$ and mass $m$, and $h$ is a dimer field with spin $1/2$ used to encode the $nc$ interactions. The spin degrees of freedom of the neutron and dimer fields are frozen, since neither the interactions nor the external current couple to spin; the problem is therefore effectively scalar. The constant $\Delta$ is the residual mass of the bare dimer field and $g$ is the strength of the $hnc$ vertex. The ellipsis denotes higher-order effective-range corrections, which are neglected.

The bare propagator of the dimer field $h$ is simply $i/\Delta$. Since the coupling $g$ is not small, $h$ is dressed by $nc$ loops to all orders. Summing the corresponding geometric series of self-energy insertions and renormalizing to reproduce a fixed pole position $\gamma$, one obtains the renormalized full propagator~\cite{Hammer:2017tjm},
\begin{equation}
\label{eq:prop-2b}
iD_{h}(p_0, \bm{p} ) = \frac{-2\pi i}{m_R g^2}
\left[-\gamma +  \sqrt{2m_R \left(\frac{\bm{p}^2}{2(M+m)}- p_0  -i\epsilon \right)} \right]^{-1},
\end{equation}
where $m_R=mM/(M+m)$ is the reduced mass of the $nc$ system. The $nc$ scattering amplitude is obtained by attaching the coupling to external $nc$ states and reads $(-ig)^2 iD_h(p_0,\bm{p})$. For positive $\gamma$, the propagator has a bound state at $E=-B=-\gamma^2/(2m_R)$ with residue 
\begin{equation}
    Z_h = \frac{2\pi\gamma}{m_R^2 g^2}\,.
\end{equation}
Thus $\gamma$ is the binding momentum of the $S$-wave halo state and $B$ is the one-neutron separation energy of the halo nucleus. The corresponding momentum-space wave function is
\begin{equation}
\psi(\bm{q})=\frac{\sqrt{8\pi\gamma}}{\gamma^2+\bm{q}^2}\,.
\end{equation}

We now consider the hard knockout of the core from the halo nucleus by an external scalar one-body current $\mathcal{J}$ coupling only to the core, and a two-body current $\mathcal{J}'$ coupling to the dimer field,
\begin{equation}
\label{eq:extcurrent}
    \mathcal{L}_{\rm ext}=-\Jc\, c^\dagger c- g^2\Jc' h^\dagger h\,,
\end{equation}
where a factor of $g^2$ has been pulled out of $\Jc'$ for convenience. For matching to the normalization used in the main text, $\Jc = f(\textbf{k}_\Jc^2)/(2M)$. We assume that both currents transfer a space-like four-momentum $k_\Jc=(\omega_\Jc,\bm{k}_\Jc)$ with $|\bm{k}_\Jc|\gg \omega_\Jc\approx \bm{k}_\Jc^2/(2M)$, which are the same assumed throughout the main body. 

The leading contribution at large momentum transfer $|\bm{k}_\Jc|$ is the plane-wave impulse approximation diagram shown in Fig.~\ref{fig:NR_1body_tree}, in which the current strikes the core. In this diagram, solid (dashed) lines indicate the neutron (core), and the wiggly line represents the external current. The spectator neutron carries four-momentum $p_{\rm 1,NR}=(\bm{p}_m^2/(2m),\bm{p}_m)$ with $|\bm{k}_\Jc|\gg|\bm{p}_m|$.

A straightforward application of the Feynman rules gives 
\begin{equation}
\Tc_\Jc =  \sqrt{Z_h}\,(-i g)\frac{i}{-B-\frac{\bm{p}_m^2}{2m}-\frac{\bm{p}_m^2}{2M}}(-i\mathcal{J}) = i \mathcal{J}\, \psi(\bm{p}_m)\,.
\end{equation}

 Note that the amplitudes computed in this appendix are non-relativistic 
in nature and should be compared with the NR amplitudes defined in 
Sec.~\ref{sec:2body_LO_NR}. Specifically, all amplitudes here correspond 
to the $1\to 2$ process, and are related to the relativistic amplitude by 
$\mathbf{T}_{1\to 2}^{\rm NR} \approx \Tc_{1\to 2}/(\sqrt{2M_b}\sqrt{2m}\sqrt{2M})$, 
cf.\ Eq.~\eqref{eq:relNR}. Rather than labeling each amplitude by its 
initial and final particle content, we instead use subscripts $\Jc$ and 
$\Jc'$ to distinguish contributions from the one-body and two-body 
currents defined in Eq.~\eqref{eq:extcurrent}. 

Now we turn our attention to kinematically suppressed contributions, starting with FSI corrections, which we label as $\Tc_\Jc^{\rm ps, FSI}$. The FSI contribution to the hard knockout from the one-body operator is shown in Fig.~\ref{fig:NR_1body_loop}, using the same notation as Fig.~\ref{fig:NR_1body_tree}. The filled circle represents the off-shell $nc$ scattering amplitude, given by $(-ig)^2 iD_h(P_{\rm i,NR}+k_\Jc)$ in our approach. Evaluating the diagram gives,
\begin{eqnarray}
\Tc_\Jc^{\rm ps, FSI} &=& \frac{\sqrt{Z_h}\, g\, \mathcal{J}\, 2\pi/m_R}{-\gamma +\sqrt{ \frac{m_R}{M+m} \bm{k}_\Jc^2-2m_R (\omega_\Jc-B)-i\epsilon}}\nonumber\\
&&\times\int\frac{d^4 q}{(2\pi)^4}\frac{1}{q_0-B-\frac{\bm{q}^2}{2m}+i\epsilon}\frac{1}{-q_0-\frac{\bm{q}^2}{2M}+i\epsilon}\frac{1}{\omega_\Jc-q_0-\frac{(\bm{k}_\Jc-\bm{q})^2}{2M}+i\epsilon}\,.
\end{eqnarray}
Performing the $q_0$ integration via the residue theorem, picking the pole at $q_0=B+\bm{q}^2/(2m)-i\epsilon$, combining the remaining denominators using the Feynman parametrization, and shifting the integration momentum $\bm{q}$ appropriately, we obtain,
\begin{equation}
\Tc_\Jc^{\rm ps, FSI} = \frac{i\, 4\pi\mathcal{J} \sqrt{8\pi\gamma}\,I}{-\gamma +\sqrt{ \frac{m_R}{M+m} \bm{k}_\Jc^2-2m_R (\omega_\Jc-B)-i\epsilon}}\,,
\end{equation}
where the finite integral $I$ is given by
\begin{equation}
I = \int_0^1 dx \int \frac{d^3 q}{(2\pi)^3}\left[ \gamma^2 +\bm{q}^2 -2xm_R \omega_\Jc  +\frac{m_R}{M}\bm{k}_\Jc^2 x\left(1-\frac{m_R}{M}x\right) -i\epsilon\right]^{-2}.
\end{equation}
For a hard knockout with $\omega_\Jc\approx \bm{k}_\Jc^2/(2M)$, a straightforward calculation yields,
\begin{eqnarray}
    I &=& \frac{M}{8\pi m_R|\bm{k}_\Jc|}
    \, \arcsin \left(\frac{m_R |\bm{k}_\Jc|}{M\gamma}\right)\nonumber\\
&=&  \frac{M}{16\pi m_R|\bm{k}_\Jc|}
    \left[\pi+2i\, \ln \left(\frac{2 m_R |\bm{k}_\Jc|}{M\gamma} \right)+ \Oc\left(\gamma^2/\bm{k}_\Jc^2\right)\right]\,.
\end{eqnarray}
Placing the final-state particles on the mass shell, i.e., setting $\omega_\Jc-B = \bm{p}_m^2/(2m)+(\bm{k}_\Jc-\bm{p}_m)^2/(2M)$, we obtain,
\begin{equation}
    \Tc_\Jc^{\rm ps, FSI} = \frac{ i\mathcal{J}\sqrt{8\pi\gamma}}{-\gamma -i\left| \bm{p}_m- \frac{m_R}{M} \bm{k}_\Jc\right|}  \frac{\pi M}{4 m_R |\bm{k}_\Jc|}\left[1+\frac{2i}{\pi}\, \ln \left(\frac{2 m_R |\bm{k}_\Jc|}{M\gamma} \right)+ \ldots\right]\,.
\end{equation}
Since $|\bm{k}_\Jc|\gg|\bm{p}_m|$, this is suppressed relative to the impulse approximation contribution $\Tc_\Jc$ by at least a factor of $\ln|\bm{k}_\Jc|/|\bm{k}_\Jc|^2$.

Finally, we consider the leading-order contribution of the two-body current $\mathcal{J}'$ in Eq.~(\ref{eq:extcurrent}) with $nc$ final-state interactions, which we label as $\Tc_{\Jc'}$. This contribution is given by the diagram on the right-hand side of Fig.~\ref{fig:bs_pole}, where the thick solid line with the filled circle represents the full dimer propagator. We obtain,
\begin{eqnarray}
\Tc_{\Jc'} &=& \frac{i\Jc' \,\pi}{m_R^2} \frac{\sqrt{8\pi\gamma}}{-\gamma -i \left| \bm{p}_m- \frac{m_R}{M} \bm{k}_\Jc\right|}\,.
\end{eqnarray}
This is a purely short-distance contribution. Since $|\bm{k}_\Jc|\gg|\bm{p}_m|$, it is suppressed relative to the impulse approximation contribution $\Tc_\Jc$ by at least one power of $1/|\bm{k}_\Jc|$. The actual size of the two-body current contribution depends on the power counting for $\Jc'$, which carries additional inverse powers of a large momentum scale compared to $\Jc$ and is thus suppressed even further. 
\end{document}